\documentclass[twocolumn]{IEEEtran}
\usepackage[T1]{fontenc}
\usepackage{amsmath}
\usepackage{amsthm}
\usepackage{amssymb}
\usepackage{graphicx}
\PassOptionsToPackage{normalem}{ulem}
\usepackage{ulem}

\makeatletter
\theoremstyle{plain}
\newtheorem{thm}{\protect\theoremname}
\theoremstyle{definition}
\newtheorem{defn}[thm]{\protect\definitionname}
\theoremstyle{plain}
\newtheorem{cor}[thm]{\protect\corollaryname}
\theoremstyle{plain}
\newtheorem{lem}[thm]{\protect\lemmaname}

\usepackage{changepage}

\IEEEoverridecommandlockouts

\newcommand{\bmat}{\left[\begin{array}}
\newcommand{\emat}{\end{array}\right]}

\makeatother

\providecommand{\corollaryname}{Corollary}
\providecommand{\definitionname}{Definition}
\providecommand{\lemmaname}{Lemma}
\providecommand{\theoremname}{Theorem}

\begin{document}

\title{Data Discovery and Anomaly Detection Using Atypicality: Signal Processing
Methods}

\author{Elyas Sabeti, \emph{Member, IEEE} and Anders H{\o}st-Madsen, \emph{Fellow,
IEEE}. \thanks{A. H{\o}st-Madsen and E. Sabeti are with the Department of Electrical
Engineering, University of Hawaii Manoa, Honolulu, HI 96822 (e-mail:
\{ahm,sabeti\}@hawaii.edu). This work was supported in part by NSF
grants CCF 1017823, 1017775, and\textbf{ }1434600 and the NSF Center
for Science of Information (CSoI). The paper was presented in part
at IEEE International Symposium on Information Theory (ISIT) 2015,
ISIT 2017, and IEEE GlobalSIP 2015.}}

\maketitle
\global\long\def\cov{\mathrm{cov}}

\begin{abstract}
The aim of atypicality is to extract small, rare, unusual and interesting
pieces out of big data. This complements statistics about typical
data to give insight into data. In order to find such ``interesting''
parts of data, universal approaches are required, since it is not
known in advance what we are looking for. We therefore base the atypicality
criterion on codelength. In a prior paper we developed the methodology
for discrete-valued data, and the the current paper extends this to
real-valued data. This is done by using minimum description length
(MDL). We show that this shares a number of theoretical properties
with the discrete-valued case. We develop the methodology for a number
of ``universal'' signal processing models, and finally apply them
to recorded hydrophone data.
\end{abstract}

\section{Introduction}

One characteristic of the information age is the exponential growth
of information, and the ready availability of this information through
networks, including the internet \textendash{} ``Big Data.'' The
question is what to do with this enormous amount of information. One
possibility is to characterize it through statistics \textendash{}
think averages. The perspective in this paper is the opposite, namely
that most of the value in the information is in the parts that deviate
from the average, that are unusual, atypical. The rest is just background
noise. Take art: the truly valuable paintings are those that are rare
and atypical. The same could be true for scientific research and entrepreneurship.
%Take online collections of photos, such as Flickr.com. Most%of the photos are rather pedestrian snapshots and not of interest%to a wider audience. The photos that of interest are those that are%unique. Flickr has a collection of photos rated for 'interestingness,'%and one can notice that those photos are indeed very different from%typical photos. They are atypical. 

The aim of our approach is to extract such 'rare interesting' data
out of big data sets. The central question is what 'interesting' means.
A first thought is to focus on the 'rare' part. That is, interesting
data is something that is unlikely based on prior knowledge of typical
data or examples of typical data, i.e., training. This is the way
an outlier is usually defined. Unlikeliness could be measured in terms
of likelihood or according to some distance measure. This is also
the most common principle in anomaly detection \cite{ChandolaAl12}.
However, perhaps being unlikely is not sufficient for something to
be 'interesting.' In many cases, outliers are junk that are eliminated
not to contaminate the typical data. What makes something interesting
is perhaps that it has a new unusual structure in itself that is quite
different than the structure of the data we have already seen. Return
to the example of paintings: what make masterworks interesting is
not just that they are different than other paintings, but that they
have some 'structure' that is intriguing. Or take another example.
Many scientific discoveries, like the theory of relativity and quantum
mechanics, began with experiments that did not fit with prevailing
theories. The experiments were outliers or anomalies. What made them
truly interesting was that it was possible to find a new theory to
explain the data, be it relativity or quantum mechanics. This is the
principle we pursue: finding data that have better alternative explanations
than those that fit the typical data.

In the paper \cite{HostSabetiWalton15} we used this intuition to
develop a methodology, atypicality, that can be used to discover such
data. The basic idea is that if some data can be encoded with a shorter
codelength in itself, i.e., with a universal source coder, rather
than using the optimum coder for typical data, then it is atypical.
The purpose of the current paper is to generalize this to real-valued
data. Lossless source coding does not generalize directly to real-valued
data. Instead we can use minimum description length (MDL). In the
current paper we develop an approach to atypicality based on MDL,
and show its usefulness on a real dataset.

\section{\label{MDL.sec}Atypicality}

We repeat the argumentation for atypicality from \cite{HostSabetiWalton15}.
Our starting point is the in theory of randomness developed by Kolmogorov
and Martin-L\"{o}f \cite{LiVitanyi,CoverBook}. Kolmogorov divides
(infinite) sequences into 'typical' and 'special.' The typical sequences
are those that we can call random, that is, they satisfy all laws
of probability. They can be characterized through Kolmogorov complexity.
A sequence of bits $\{x_{n},n=1,\ldots,\infty\}$ is random (i.e,
iid uniform) if the Kolmogorov complexity of the sequence satisfies
$K(x_{1},\ldots,x_{n})\geq n-c$ for some constant $c$ and for all
$n$ \cite{LiVitanyi}. The sequence is incompressible if $K(x_{1},\ldots,x_{n}|n)\geq n$
for all $n$, and a finite sequence is algorithmically random if $K(x_{1},\ldots,x_{n}|n)\geq n$
\cite{CoverBook}. In terms of coding, an iid random sequence is also
incompressible, or, put another way, the best coder is the identity
function. Let us assume we draw sequences $x^{n}$ from an iid uniform
distribution. The optimum coder is the identity function, and the
code length is $n$. Now suppose that for one of these sequences we
can find a (universal) coder so that the code length is less than
$n$; while not directly equivalent, one could state this as $K(x_{1},\ldots,x_{n}|n)<n$.
With an interpretation of Kolmogorov's terms, this would not be a
'typical' sequence, but a 'special' sequence. We will instead call
such sequences 'atypical.' Considering general distributions and general
(finite) alphabets instead of iid uniform distributions, we can state
this in the following general principle \cite{HostSabetiWalton15}
\begin{defn}
\label{atypdef.thm}\emph{A sequence is atypical if it can be described
(coded) with fewer bits in itself rather than using the (optimum)
code for typical sequences}.
\end{defn}

\section{\label{Real.sec}Real-valued models}

We would like to extend definition \ref{atypdef.thm} and the approach
in \cite{HostSabetiWalton15} to real-valued models. The approach
in \cite{HostSabetiWalton15} can at a high level be described as
comparing a typical coder based on fixed codes (e.g., Huffman codes
for given probabilities) with an atypical coder based on a universal
source coder; if the universal code is shorter, the sequence is declared
atypical. As in \cite{HostSabetiWalton15} we would like to locate
fixed length sequences, variable length sequences, and subsequences
of variable length.

The definition is based on \emph{exact} description of data, and lossless
source coding rather than lossy (rate-distortion) therefore is the
appropriate generalization. Lossless coding of real-valued data is
used in many applications, for example lossless audio coding \cite{GhidoTabus13}.
Direct encoding of the reals represented as binary numbers, such as
done in lossless audio coding, makes the methods too dependent on
data representation rather than the underlying data. Instead we will
use a more abstract model of (finite-precision) reals. We will assume
a fixed point representation with a (large) finite number, $r$, bits
after the period, and an \emph{unlimited} number of bits prior to
the period \cite{Rissanen83}. Assume that the actual data is distributed
according to a pdf $f(x)$. Then the number of bits required to represent
$x$ is given by
\begin{align}
L(x) & =-\log\int_{x}^{x+2^{-r}}f(t)dt\approx-\log(f(x)2^{-r})\nonumber \\
 & =-\log(f(x))+r\label{RealLength.eq}
\end{align}
As we are only interested in \emph{comparing} codelengths the dependency
on $r$ cancels out. Suppose we want to decide between two models
$f_{1}(x)$ and $f_{2}(x)$ for data. Then we decide $f_{1}(x)$ if
$\lim_{r\to\infty}-\log\int_{x}^{x+2^{-r}}f_{1}(t)dt+\log\int_{x}^{x+2^{-r}}f_{2}(t)dt>0$,
which is $-\log f_{1}(x)>-\log f_{2}(x)$. Thus for the \emph{typical}
\emph{codelength} we can simply use $L_{t}(x)=-\log f(x)$, where
$f(x)$ is the \emph{known} distribution of typical data. One can
also argue for this codelength more fundamentally from finite blocklength
rate-distortion in the limit of low distortion \cite{Kostina17}.
Notice that this codelength is not scaling invariant:
\begin{align}
y & =ax+b\nonumber \\
L_{t}(y) & =-\log f(x)+\log|a|\label{scaling.eq}
\end{align}
which means care has to be taken when transforms of data are considered. 

For the atypical codelength, there is nothing like universal source
coding for the reals. The principle of universal source coding is
that the transmitted sequence allows decoding of both the sequence
and potential unknown parameters (this is the idea in the CTW algorithm
\cite{WillemsAl95} used in \cite{HostSabetiWalton15}). That principle
is similar to that used in Rissanen's minimum description length (MDL)
\cite{Rissanen83}. The MDL is generally a codelength based on a specific
model; on the other hand, in atypicality we are not interested in
finding out if the data follows a specific model. Therefore, our approach
is to try to code the data with a set of various \emph{general} data
(signal processing) models hoping that one of them approximates the
actual model of the data better than the typical model. Let the models
be $\mathcal{M}_{i,k}$, where the first index denotes the model type
and the second index the number of (real) parameters. The atypical
codelength for a sequence $x^{l}=\{x_{1},x_{2},\ldots,x_{l}\}$ then
is 
\begin{equation}
L_{a}(x^{l})=\min_{i,k}L(x^{l}|\mathcal{M}_{i,k})+L(i,k)\label{mix.eq}
\end{equation}
where $L(x^{l}|\mathcal{M}_{i,k})$ is the codelength to encode $x^{l}$
with the model $\mathcal{M}_{i,k}$ \emph{including any parameters}
and $L(i,k)$ is some code to tell the decoder which model is used.
An even better approach is to use weighting as in \cite{WillemsAl95},
\[
L_{a}(x^{l})=-\log\left(\sum_{i,k}w_{i,k}2^{-L(x^{l}|\mathcal{M}_{i,k})}\right),
\]
where $\sum_{i,k}w_{i,k}\leq1$. A central tenet of atypicality is
an adherence to strict decodability: we imagine that there is a receiver
that receives solely a sequence of bits, and from this it should be
able to reconstruct the data. Thus, the codelengths $L(x^{l}|\mathcal{M}_{i,k})$
should be actual lengths. On the other hand, strict universality (as
in for example normalized maximum likelihood \cite{GrunwaldBook})
is less central. Even if each code $L(x^{l}|\mathcal{M}_{i,k})$ satisfies
some strict universality criterion, the final codelength $L_{a}(x^{l})$
might not necessarily satisfy this. So, universality in some vague
sense is sufficient.

The most common application of MDL is model selection and choosing
number of parameters in the model. Thus, two codelengths $L(x^{l}|\mathcal{M}_{i,k})$
and $L(x^{l}|\mathcal{M}_{\tilde{i},\tilde{k}})$ are compared. In
atypicality, on the other hand, the MDL codelength is only compared
to the typical codelength $L_{t}(x^{l})$. This has a number of consequences.
First, we might use different MDL methods for different models \textendash{}
this makes little sense in model selection, but perfect sense in atypicality.
Second, we might use different models for different parts of the sequence
$x^{l}$. Finally, again strict universality is less important. 

In atypicality, as mentioned at the start of the section, we are also
interested in finding sequences of variable length. It is therefore
important that any MDL principle used works for both short and long
sequences.

\section{Minimum Description Length (MDL)}

Let $f(x^{l}|\boldsymbol{\theta})$ denote a pdf for the sequence
$x^{l}$ parametrized the $k$-dimensional parameter vector $\boldsymbol{\theta}$.
Rissanen's famous MDL approach \cite{Rissanen83} is a way to jointly
encode the sequence $x^{l}$ and the unknown parameters $\boldsymbol{\theta}$.
A widely known expression for codelength, frequently used in signal
processing, is
\begin{equation}
L=-\log f(x^{l}|\hat{\boldsymbol{\theta}})+\frac{k}{2}\log l\label{RealMDL.eq}
\end{equation}
where $\hat{\boldsymbol{\theta}}$ is the maximum likelihood (ML)
estimate. The expression (\ref{RealMDL.eq}) is known to be a quite
good approximation for many actual MDL coding methods \cite{GrunwaldBook},
e.g., within an $O(1)$ term under some restrictive assumptions.

One possible approach to generalizing atypicality to real-valued data
is therefore to simply use the expression (\ref{RealMDL.eq}) as the
atypical codelength. This has the advantage that we can easily take
any signal processing model, count the number of unknown parameters,
and then use (\ref{RealMDL.eq}); we believe this is a valid approach
to atypicality. It also had the advantage that it is possible to derive
analytical results, see Section \ref{asymptotic.sec}.

However, the issue is still that (\ref{RealMDL.eq}) is not a true
codelength. As mentioned in Section \ref{Real.sec}, a tenet of atypicality
is to use actual codelength; additionally, we would like to analyze
sequences of variable length $l$, where perhaps even $k$ is increasing
with $l$. We would also like to apply atypicality to mixed type data
that has both discrete and real components. The discrete coder returns
an actual codelength, so that is better combined with an actual real-valued
codelength\footnote{the term $r$ in (\ref{RealLength.eq}) still cancels out in comparison,
so the fact that the discrete codelength is finite while the real-valued
codelength is infinite is not an issue.}. More generally, when combining multiple methods as in (\ref{mix.eq}),
the arbitrary constant in (\ref{RealMDL.eq}) (or terms of order less
than $\log l$) influences detection and false alarm probabilities
critically, and this gives issues with using (\ref{RealMDL.eq}).
Finally (\ref{RealMDL.eq}) is difficult to directly apply when using
transform-coding as in Sections \ref{Filterbank.sec} and \ref{subsec:DFT}.

For use in atypicality we therefore introduce two new MDL methods,
based on a common principle. Our starting point is Rissanen's \cite{Rissanen86}
original predictive MDL 
\begin{equation}
L(x^{l})=-\sum_{i=0}^{l-1}\log f(x_{i+1}|\boldsymbol{\hat{\theta}}(x^{i}))\label{predictiveMDL.eq}
\end{equation}
The issue with this method is how to initialize the recursion. When
$i=0$, $\boldsymbol{\hat{\theta}}(x^{i})$ is not defined. Rissanen
suggests using a default pdf $f_{d}$ to encode data until $\boldsymbol{\hat{\theta}}(x^{i})$
is defined, so that $L(x^{l})=-\sum_{i=1}^{l-1}\log f(x_{i+1}|\boldsymbol{\hat{\theta}}(x^{i}))-\log f_{d}(x_{1})$.
In general, with more than one parameter, the default pdf might have
to be used for more samples. The remaining issue is that even when
$\boldsymbol{\hat{\theta}}(x^{i})$ \emph{is} defined, the estimate
might be poor, and using this in (\ref{predictiveMDL.eq}) can give
very long codelengths, see Fig. \ref{fig:Redundancy} below. Our solution
is rather than using the ML estimate for encoding as though it is
the \emph{actual} parameter value, we use it as an \emph{uncertain}
estimate of $\boldsymbol{\theta}$. We then take this uncertainty
into account in the codelength. This is similar to the idea of using
confidence intervals in statistical estimates \cite{larsen1986introduction}.
Below we introduce two methods using this general principle. This
is different than the sequentially normalized maximum likelihood method
\cite{RoosRissanen08}, which modifies the encoder itself.

\subsection{Subsequences}

As in \cite{HostSabetiWalton15} our main interest is to find atypical
subsequences of long sequences. The main additional consideration
here is that when an atypical subsequence is encoded, the decoder
also needs to know the start and end of the sequence. As described
in \cite{HostSabetiWalton15}, the start is encoded with a special
codeword of length $\tau$ bits, where $\tau\approx-\log P(\text{\text{'atypical'})}$,
and the end is encoded by transmitting the length of the sequence,
which \cite{Rissanen83,Elias75} can be done with $\log^{*}l+\log c$,
where $c$ is a constant and $\log^{*}(l)=\log l+\log\log l+\log\log\log l+\cdots$.
If we use (\ref{RealMDL.eq}) only the first term matters, and we
get a subsequence codelength
\begin{equation}
L=-\log f(x^{l}|\hat{\boldsymbol{\theta}})+\frac{k+2}{2}\log l+\tau\label{RealMDLsub.eq}
\end{equation}
In principle the term $\tau$ does not matter as there are unknown
constants, but $\tau$ is useful as a threshold. As shown in \cite{HostSabetiWalton15}
the extra $\log l$ term is essential to obtain a finite atypical
subsequence probability. In the following we will principally consider
the subsequence problem.

\subsection{\label{asymptotic.sec}Asymptotic MDL}

In this section we assume (\ref{RealMDLsub.eq}) is used as codelength.
Developing algorithms is straightforward: we just use various maximum
likelihood estimators and count the number of parameters. Examples
can be found in \cite{Host15ISIT}. Here we will focus on performance
analysis.

Consider a simple example. The typical model is a pure zero-mean Gaussian
noise model with known variance $\sigma^{2}$. For the atypical model
we let $x\sim\mathcal{N}(\mu_{a},\sigma^{2}$) with $\mu_{a}$ unknown.
The typical codelength is
\begin{align*}
L_{t}(l) & =-\sum\log\left(\frac{1}{\sqrt{2\pi\sigma^{2}}}\exp\left(-\frac{x[n]^{2}}{2\sigma^{2}}\right)\right)\\
 & =\frac{l}{2}\log2\pi\sigma^{2}+\sum\frac{x[n]^{2}}{2\sigma^{2}\ln2}
\end{align*}
The ML estimate of the one unknown parameter $\mu_{a}$ is the average
$\bar{x}$, and we get a codelength using (\ref{RealMDLsub.eq})
\begin{align*}
L_{a}(l) & =\frac{l}{2}\log2\pi\sigma^{2}+\sum\frac{\left(x[n]-\bar{x}\right)^{2}}{2\sigma^{2}\ln2}+\frac{3}{2}\log l+\tau
\end{align*}
The criterion for atypicality is
\[
L_{t}(l)-L_{a}(l)=\frac{l\bar{x}^{2}}{2\sigma^{2}\ln2}-\frac{3}{2}\log l-\tau>0
\]
or
\[
\left|\frac{1}{\sqrt{l}}\sum x[n]\right|>\sigma\sqrt{3\ln l+(2\tau+5)\ln2}
\]
If the data is typical, $\frac{1}{\sqrt{l}}\sum x[n]\sim\mathcal{N}(0,\sigma^{2})$.
Of key theoretical interest is the probability that a sequence generated
according to the typical model is classified is atypical. One can
think of this as a false alarm, but since the sequence is indistinguishable
from one generated from an alternative model, we prefer the term \emph{\uline{intrinsically
atypical}} \cite{HostSabetiWalton15}.

The probability of a sequence being intrinsically atypical is upper
bounded by \cite{VerduBook}
\begin{align*}
P_{A}(l) & =2Q\left(\sqrt{3\ln l+(2\tau+5)\ln2}\right)\\
 & \leq\exp\left(-(3\ln l+(2\tau+5)\ln2)/2\right)\\
 & =2^{-5/2}l^{-3/2}2^{-\tau}.
\end{align*}
and lower bounded by
\begin{align*}
P_{A}(l) & >\frac{2}{\sqrt{2\pi(3\ln l+(2\tau+5)\ln2)}}\left(1-\frac{1}{3\ln l+(2\tau+5)\ln2}\right)\\
 & \times\exp\left(-(3\ln l+(2\tau+5)\ln2)/2\right)
\end{align*}
from which we conclude
\begin{equation}
\lim_{l\to\infty}\frac{\ln P_{A}(l)}{-\frac{3}{2}\ln l}=1\label{MeanDecay.eq}
\end{equation}
It is interesting that this is the same expression (except for constant
factors) as for the iid binary case in \cite{HostSabetiWalton15}.
It means that, using the Gaussian mean criterion is equivalent to
using the binary criterion on the \emph{sign} of the samples. This
illustrates that the discrete version of atypicality and real-valued
version are part of one unified theory. 

For the general vector Gaussian case, we have the following result
\begin{thm}
\label{kupper.thm}Suppose that the typical model is $\mathcal{N}(\mathbf{s},\boldsymbol{\Sigma})$
and the atypical model is $\mathcal{N}(\mathbf{s}(\boldsymbol{\theta}),\boldsymbol{\Sigma}(\boldsymbol{\theta}))$,
where $\boldsymbol{\theta}$ is $k$-dimensional. Then the probability
$P_{A}(l)$ of an intrinsically atypical subsequence is bounded by
\begin{equation}
\limsup_{l\to\infty}\frac{\ln P_{A}(l)}{-\frac{k+2}{2}\ln l}\leq1\label{Vbound.eq}
\end{equation}
\end{thm}
\begin{IEEEproof}
For simplicity of notation, in this proof we will assume codelength
is in nats and use natural logarithms throughout. We can precode the
data with the typical model, so that after precoding we can assume
the typical model is $\mathcal{N}(0,\mathbf{I}).$ The atypicality
criterion is
\[
r(\mathbf{x})=-\ln\frac{f(\mathbf{x}|\hat{\boldsymbol{\theta})}}{f(\mathbf{x})}\geq\tau+\frac{k+2}{2}\ln l
\]
The Chernoff bound now states that for any $s>0$
\begin{align*}
P\left(r(\mathbf{x})\geq\tau+\frac{k+2}{2}\ln l\right) & \leq\exp(-s(\tau+\frac{k+2}{2}\ln l))M_{r}(s)
\end{align*}
where $M_{r}(s)=E[e^{sr}]$. If we put $s=1$ we obtain (\ref{Vbound.eq}),
provided $M_{r}(s)$ is bounded as $l\to\infty$. We will prove that
$M_{r}(s)\leq K<\infty$ independent of $l$ for any $s<1$, which
is sufficient to state (\ref{Vbound.eq}) by letting $s\to1$ sufficiently
slow as $l\to\infty$.

We have
\begin{align*}
-\ln\frac{f(\mathbf{x}|\hat{\boldsymbol{\theta})}}{f(\mathbf{x})} & =\frac{1}{2}\sum_{n=1}^{l}\mathbf{x}_{n}^{T}\mathbf{x}_{n}\\
 & -\frac{1}{2}\sum_{n=1}^{l}\left(\mathbf{x}_{n}-\mathbf{s}(\hat{\boldsymbol{\theta}})\right)^{T}\boldsymbol{\hat{\Sigma}}(\boldsymbol{\theta})^{-1}\left(\mathbf{x}_{n}-\mathbf{s}(\hat{\boldsymbol{\theta}})\right)\\
 & -\frac{l}{2}\ln\det\boldsymbol{\hat{\Sigma}}(\boldsymbol{\theta})
\end{align*}
We need to upper bound this expression. Maximum likelihood estimation
is given by minimizing the second and third terms over all $(\mathbf{s}(\hat{\boldsymbol{\theta}}),\boldsymbol{\hat{\Sigma}}(\boldsymbol{\theta})),\boldsymbol{\theta}\in\mathbb{R}^{k}$.
The set $(\mathbf{s}(\hat{\boldsymbol{\theta}}),\boldsymbol{\hat{\Sigma}}(\boldsymbol{\theta})),\boldsymbol{\theta}\in\mathbb{R}^{k}$
is a manifold in $\mathbb{R}^{M}\times\mathbb{R}^{M^{2}}$. Minimizing
over all (valid) vectors $\mathbb{R}^{M}\times\mathbb{R}^{M^{2}}\in\mathbb{R}^{M}$
can only make the term smaller, and the minimizer is of course the
ML estimate, here $\mathbf{y}=(\hat{\boldsymbol{\mu}},\hat{\boldsymbol{\Sigma}})$
with $\hat{\boldsymbol{\mu}}=\frac{1}{l}\sum_{i=1}^{l}\mathbf{x}_{i}$,
$\hat{\boldsymbol{\Sigma}}=\frac{1}{l}\sum_{i-1}^{l}(\mathbf{x}_{i}-\hat{\boldsymbol{\mu}})(\mathbf{x}_{i}-\hat{\boldsymbol{\mu}})^{T}$.
Thus
\begin{align*}
-\ln\frac{f(\mathbf{x}|\hat{\boldsymbol{\theta})}}{f(\mathbf{x})} & \leq\frac{1}{2}\sum_{n=1}^{l}\mathbf{x}_{n}^{T}\mathbf{x}_{n}\\
 & -\frac{1}{2}\sum_{n=1}^{l}\left(\mathbf{x}_{n}-\hat{\boldsymbol{\mu}}\right)^{T}\hat{\boldsymbol{\Sigma}}^{-1}\left(\mathbf{x}_{n}-\hat{\boldsymbol{\mu}}\right)-\frac{l}{2}\ln\det\boldsymbol{\hat{\Sigma}}\\
 & =\frac{1}{2}\sum_{n=1}^{l}\mathbf{x}_{n}^{T}\mathbf{x}_{n}-\frac{lM}{2}-\frac{l}{2}\ln\det\boldsymbol{\hat{\Sigma}}\\
 & =\frac{l}{2}\text{tr}\boldsymbol{\hat{\Sigma}}-\frac{lM}{2}-\frac{l}{2}\ln\det\boldsymbol{\hat{\Sigma}}+\frac{l}{2}\hat{\boldsymbol{\mu}}^{T}\hat{\boldsymbol{\mu}}
\end{align*}
and
\begin{align}
E[e^{sr}] & \leq E\left[\exp\left(\frac{sl}{2}\text{tr}\boldsymbol{\hat{\Sigma}}-\frac{slM}{2}-\frac{sl}{2}\ln\det\boldsymbol{\hat{\Sigma}}\right)\exp\left(\frac{sl}{2}\hat{\boldsymbol{\mu}}^{T}\hat{\boldsymbol{\mu}}\right)\right]\nonumber \\
 & \leq E\left[\exp\left(\frac{sl}{2}\text{tr}\boldsymbol{\hat{\Sigma}}-\frac{slM}{2}-\frac{sl}{2}\ln\det\boldsymbol{\hat{\Sigma}}\right)\right]\nonumber \\
 & \times E\left[\exp\left(\frac{sl}{2}\hat{\boldsymbol{\mu}}^{T}\hat{\boldsymbol{\mu}}\right)\right]\label{Esrbound.eq}
\end{align}
For the latter expectation we use that $\hat{\boldsymbol{\mu}}\sim\mathcal{N}(0,\frac{1}{l}\mathbf{I})$.
We can therefore write

\begin{align*}
E\left[\exp\left(\frac{sl}{2}\hat{\boldsymbol{\mu}}^{T}\hat{\boldsymbol{\mu}}\right)\right] & \leq\frac{\sqrt{l}}{(2\pi)^{l/2}}\int\exp\left(\frac{sl}{2}\mathbf{t}^{T}\mathbf{t}\right)\exp\left(-\frac{l}{2}\mathbf{t}^{T}\mathbf{t}\right)d\mathbf{t}\\
 & \leq K
\end{align*}
for $s<1$.

We rewrite the first expectation in (\ref{Esrbound.eq}) as
\begin{align*}
\lefteqn{E\left[\exp\left(\frac{sl}{2}\text{tr}\boldsymbol{\hat{\Sigma}}-\frac{slM}{2}-\frac{sl}{2}\ln\det\boldsymbol{\hat{\Sigma}}\right)\right]}\\
 & =E\left[\exp\left(\frac{sl}{2}\text{tr}\left(\frac{(l-1)}{(l-1)}\boldsymbol{\hat{\Sigma}}\right)-\frac{slM}{2}-\frac{sl}{2}\ln\det\left(\frac{(l-1)}{(l-1)}\boldsymbol{\hat{\Sigma}}\right)\right)\right]\\
 & =E\left[\exp\left(\frac{sl}{2(l-1)}\text{tr}\boldsymbol{\Sigma}-\frac{slM}{2}-\frac{sl}{2}\ln\left(\frac{1}{(l-1)^{M}}\det\boldsymbol{\Sigma}\right)\right)\right]\\
 & =E\left[\exp\left(\frac{sl}{2(l-1)}\text{tr}\boldsymbol{\Sigma}-\frac{slM}{2}+\frac{slM}{2}\ln\left(l-1\right)-\frac{sl}{2}\ln\det\boldsymbol{\Sigma}\right)\right]
\end{align*}
Here $\boldsymbol{\Sigma}=(l-1)\hat{\boldsymbol{\Sigma}}$ , which
is known to have a Wishart distribution $\mathcal{W}_{M}(\mathbf{I},l-1)$
\cite{muirhead2009MultivariateStat} with pdf 

\[
f(\boldsymbol{\Sigma})=\frac{1}{2^{(l-1)M/2}\Gamma_{M}(\frac{l-1}{2})}\left(\det\boldsymbol{\Sigma}\right)^{(l-M-2)/2}\exp\left(-\frac{1}{2}\text{tr}\boldsymbol{\Sigma}\right)
\]
The expectation can now be evaluated as the integral
\begin{align*}
I & =\alpha\int_{\boldsymbol{\Sigma}>0}\exp\left(\frac{s\frac{l}{l-1}-1}{2}\text{tr}\boldsymbol{\Sigma}\right)\left(\det\boldsymbol{\Sigma}\right)^{((1-s)l-M-2)/2}d\boldsymbol{\Sigma}\\
 & =\alpha\Gamma_{M}\left(\frac{(1-s)l-1}{2}\right)\left(\frac{s\frac{l}{l-1}-1}{2}\right)^{-M((1-s)l-M-2)/2-1}
\end{align*}
where $\alpha$ is a factor independent of $\boldsymbol{\Sigma}$
\begin{align*}
\alpha & =\frac{\left(l-1\right)^{\frac{slM}{2}}\exp\left(-\frac{slM}{2}\right)}{2^{(l-1)M/2}\Gamma_{M}(\frac{l-1}{2})}
\end{align*}
and $\Gamma_{M}$ is the multivariate gamma function \cite{muirhead2009MultivariateStat}.
Using Stirling's approximation repeatedly, and performing some lengthy
but straightforward simplifications we then get
\begin{align*}
I & \sim\left(\frac{(1-s)l-1}{l-1}\right)^{M(1-M)/2}\\
 & \leq K
\end{align*}
when $s<1$.

\end{IEEEproof}
\begin{cor}
\label{gupper.thm}Suppose that we consider a finite set of atypical
signal models $\{\mathbf{s}(\boldsymbol{\theta}),\boldsymbol{\Sigma}(\boldsymbol{\theta})\}$.
Then 
\[
\limsup_{l\to\infty}\frac{\ln P_{A}(l)}{-\frac{3}{2}\ln l}\leq1
\]
\end{cor}
\begin{IEEEproof}
We can use the union bound over the different models. The models with
slowest decay in $l$ will dominate for large $l$, and these are
exactly the one-parameter models.
\end{IEEEproof}
On the other hand, we know from (\ref{MeanDecay.eq}) that for the
simple mean, the probability of an atypical sequence is exactly $\sim l^{-3/2}$.
Thus, adding more complex models will not change this by the Corollary.
This is the benefit of using MDL: searching over very complex models
will not increase the probability of intrinsically atypical sequences,
or in terms of anomaly detection, the false alarm probability.

\subsection{Normalized Likelihood Method (NLM)}

\label{subsec:Improper-Uniform-Prior}

As explained previously, our approach to predictive MDL is to introduce
uncertainty in the estimate of $\boldsymbol{\theta}$. The first method
is very simple. Let the likelihood function of the model be $f(x^{l}|\boldsymbol{\theta})$.
For a fixed $x^{l}$ we can consider this as a ``distribution''
on $\boldsymbol{\theta}$; the ML estimate is of course the most likely
value of this distribution. To account for uncertainty in the estimate,
we can instead try use the total $f(x^{l}|\boldsymbol{\theta})$ to
give a \emph{distribution} on $\boldsymbol{\theta}$, and then use
this for prediction. In general $f(x^{l}|\boldsymbol{\theta})$ is
not a probability distribution as it does not integrate to 1 in $\boldsymbol{\theta}$.
We can therefore normalize it to get a probability distribution 
\begin{equation}
f_{x^{l}}(\boldsymbol{\theta})=\frac{f(x^{l}|\boldsymbol{\theta})}{C(x^{l})};\quad C(x^{l})=\int f(x^{l}|\boldsymbol{\theta})d\boldsymbol{\theta}\label{NormLikelihood.eq-1}
\end{equation}
if $\int f(x^{l};\boldsymbol{\theta})d\boldsymbol{\theta}$ is finite.
For comparison, the Bayes posteriori distribution is 
\[
f(\boldsymbol{\theta}|x^{l})=\frac{f(x^{l}|\boldsymbol{\theta})f(\boldsymbol{\theta)}}{\int f(x^{l}|\boldsymbol{\theta})f(\boldsymbol{\theta)}d\boldsymbol{\theta}}
\]
If the support $\Theta$ of $\boldsymbol{\theta}$ has finite area,
(\ref{NormLikelihood.eq-1}) is just the Bayes predictor with uniform
prior. If the support $\Theta$ of $\boldsymbol{\theta}$ does not
have finite area, we can get (\ref{NormLikelihood.eq-1}) as a limiting
case when we take the limit of uniform distributions on finite $\Theta_{n}$
that converge towards $\Theta$. This is the same way the ML estimator
can be seen as a MAP estimator with uniform prior \cite{ScharfBook}.
One can reasonably argue that if we have no further information about
$\boldsymbol{\theta}$, a uniform distribution seems reasonable, and
has indeed been used for MDL \cite{GrunwaldBook} as well as universal
source coding \cite[Section 13.2]{CoverBook}. What the Normalized
Likelihood Method does is simply extend this to the case when there
is no proper uniform prior for $\boldsymbol{\theta}$. 

The method was actually implicitly mentioned as a remark by Rissanen
in \cite[Section 3.2]{rissanen1998stochastic}, but to our knowledge
was never further developed; the main contribution in this paper is
to introduce the method as a practical method. From Rissanen we also
know the coding distribution for $x_{n}$ 
\begin{equation}
f(x_{n+1}|x^{n})=\int f(x_{n+1}|\boldsymbol{\theta})f_{x^{n}}(\boldsymbol{\theta})d\boldsymbol{\theta}=\frac{C\left(x^{n+1}\right)}{C\left(x^{n}\right)}\label{Cexp.eq}
\end{equation}
Let us assume $C(x^{n})$ becomes finite for $n>1$ (this is not always
the case, often $n$ needs to be larger). The total codelength can
then be written as 
\begin{align}
L(x^{l}) & =\sum_{i=1}^{l-1}-\log f(x_{i+1}|x^{i})-\log f_{d}(x_{1})\nonumber \\
 & =-\log C(x^{l})+\log C(x^{2})-\log f_{d}(x_{1})\label{L1.eq}
\end{align}

\subsection{Sufficient Statistic Method (SSM)}

\label{subsec:Method-of-Sufficient}The second method for introducing
uncertainty in the estimate of $\boldsymbol{\theta}$ is more intricate.
It is best explained through a simple example. Suppose our model is
$\mathcal{N}(\mu,\sigma^{2})$, with $\sigma$ known. The average
$\bar{x}_{n}$ is the ML estimate of $\mu$ at time $n$. We know
that 
\[
\bar{x}_{n}=\mu+z,\quad z\sim\mathcal{N}\left(0,\frac{\sigma^{2}}{n}\right).
\]
We can re-arrange this as
\[
\mu=\bar{x}_{n}-z
\]
Thus, \emph{given} $\bar{x}_{n}$, we can think of $\mu$ as random
$\mathcal{N}\left(\bar{x}_{n},\frac{\sigma^{2}}{n}\right)$. Now 
\[
x_{n+1}=\mu+z_{n+1}\sim\mathcal{N}\left(\bar{x}_{n},\sigma^{2}+\frac{\sigma^{2}}{n}\right)
\]
which we can use as a coding distribution for $x_{n+1}$. This compares
to $\mathcal{N}\left(\bar{x}_{n},\sigma^{2}\right)$ that we would
use in traditional predictive MDL. Thus, we have taken into account
that the estimate of $\mu$ is uncertain for $n$ small. The idea
of thinking of the non-random parameter $\mu$ as random is very similar
to the philosophical argument for confidence intervals \cite{larsen1986introduction}.

In order to generalize this example to more complex models, we take
the following approach. Suppose $\mathbf{t}(x^{n})$ is a $k$-dimensional
sufficient statistic for the $k$-dimensional $\boldsymbol{\theta}\in\Theta$.
Also suppose there exists some function $\mathbf{s}$ and a $k$-dimensional
(vector) random variable $\mathbf{Y}$ \emph{independent of $\boldsymbol{\theta}$}
so that 
\begin{equation}
\mathbf{t}(x^{n})=\mathbf{s}(\mathbf{Y},\boldsymbol{\theta}).\label{tf.eq}
\end{equation}
We now assume that for every $(\mathbf{t},\mathbf{Y})$ in their respective
support (\ref{tf.eq}) has a solution for $\boldsymbol{\theta}\in\Theta$
so that we can write
\begin{equation}
\boldsymbol{\theta}=\mathbf{r}(\mathbf{Y},\mathbf{t}(x^{n})).\label{trY.eq}
\end{equation}
The parameter $\boldsymbol{\theta}$ is now a random variable (assuming
$\mathbf{r}$ is measurable, clearly) with a pdf $f_{x^{n}}(\boldsymbol{\theta})$
This then gives a distribution on $x_{n+1}$, i.e., 
\begin{align}
f(x_{n+1}|x^{n}) & =\int f(x_{n+1}|\boldsymbol{\theta})f_{x^{n}}(\boldsymbol{\theta})d\boldsymbol{\theta}\label{fxn1SSM.eq}
\end{align}

The method has the following property 
\begin{thm}
\label{Invariant.thm}The distribution of $x_{n+1}$ is invariant
to arbitrary parameter transformations. 
\end{thm}
This is a simple observation from the fact that (\ref{fxn1SSM.eq})
is an expectation, and that when $\boldsymbol{\theta}$ is transformed,
the distribution according to (\ref{trY.eq}) is also transformed
with the same function.

One concern is the way the method is described. Perhaps we could use
different functions $\mathbf{s}$ and $\mathbf{r}$ and get a different
result? In the following we will prove that the distribution of $\boldsymbol{\theta}$
is independent of which $\mathbf{s}$ and $\mathbf{r}$ are used. 

It is well-known \cite{GrimmettBook,CoverBook} that if the random
variable $X$ has CDF $F$, then $U=F(X)$ has a uniform distribution
(on $[0,1]$). Equivalently, $X=F^{-1}(U)$ for some uniform random
variable $U$. We need to generalize this to $n$ dimensions. Recall
that for a continuous random variable \cite{GrimmettBook} 
\begin{align*}
\lefteqn{F_{i|i-1,\ldots,1}(x_{i}|x_{i-1},\ldots x_{1})=\int_{-\infty}^{x_{i}}f(t|x_{i-1},\ldots,x_{1})dt}\\
 & =\frac{1}{f(x_{i-1},\ldots,x_{1})}\int_{-\infty}^{x_{i}}f(t,x_{i-1},\ldots,x_{1})dt
\end{align*}
whenever $f(x_{i-1},\ldots,x_{1})\neq0$. As an example, let $n=2$.
Then the map $(X_{1},X_{2})\mapsto(F_{1}(X_{1}),F_{2|1}(X_{2},X_{1}))$
is a map from $\mathbb{R}^{2}$ onto $[0,1]^{2}$, and $(F_{1}(X_{1}),F_{2|1}(X_{2},X_{1}))$
has uniform distribution on $[0,1]^{2}$. Here $F_{1}(X_{1})$ is
continuous in $X_{1}$ and $F_{2|1}(X_{2},X_{1})$ is continuous in
$X_{2}$

We can write $X_{1}=F_{1}^{-1}(U_{1})$. For fixed $x_{1}$ we can
also write $X_{2}=F_{2|1}^{-1}(U_{2}|x_{1})$ for those $x_{1}$ where
$F_{2|1}$ is defined, and where the inverse function is only with
respect to the parameter before $|$. Then
\[
\left[\begin{array}{c}
X_{1}\\
X_{2}
\end{array}\right]=\left[\begin{array}{c}
F_{1}^{-1}(U_{1})\\
F_{2|1}^{-1}(U_{2}|F_{1}^{-1}(U_{1}))
\end{array}\right]\triangleq\check{\mathbf{F}}^{-1}(U_{1},U_{2})
\]
This gives the correct joint distribution on $(X_{1},X_{2})$: the
marginal distribution on $X_{1}$ is correct, and the conditional
distribution of $X_{2}$ given $X_{1}$ is also correct, and this
is sufficient. Clearly $\check{\mathbf{F}}^{-1}$ is not defined for
all $U_{1},U_{2}$; the relationship should be understood as being
valid for almost all $(X_{1},X_{2})$ and $(U_{1},U_{2})$. We can
now continue like this for $X_{3},X_{4},\ldots,X_{n}$. We will state
this result as a lemma
\begin{lem}
\label{Generate.thm}For any continuous random variable $\mathbf{X}$
there exists an $n$-dimensional uniform random variable $\mathbf{U}$,
so that $\mathbf{X}=\check{\mathbf{F}}^{-1}(\mathbf{U})$.
\end{lem}

\begin{thm}
\label{thm:Uniqueness}Consider a model $\mathbf{t}=\mathbf{s}_{1}(\mathbf{Y}_{1};\boldsymbol{\theta}),$
with $\boldsymbol{\theta}=\mathbf{r}_{1}(\mathbf{Y}_{1};\mathbf{t})$
and an alternative model $\mathbf{t}=\mathbf{s}_{2}(\mathbf{Y}_{2};\boldsymbol{\theta}),$
with $\boldsymbol{\theta}=\mathbf{r}_{2}(\mathbf{Y}_{2};\mathbf{t}).$
We make the following assumptions
\begin{enumerate}
\item \label{connected.enu}The support of $\mathbf{t}$ is independent
of $\boldsymbol{\theta}$ and its interior is connected.
\item The extended CDF $\check{\mathbf{F}}_{i}$ of $\mathbf{Y}_{i}$ is
continuous and differentiable.
\item \label{assumps.enu}The function $\mathbf{Y}_{i}\mapsto\mathbf{s}_{i}(\mathbf{Y}_{i};\boldsymbol{\theta})$
is one-to-one, continuous, and differentiable for fixed $\boldsymbol{\theta}$.
\end{enumerate}
Then the distributions of $\boldsymbol{\theta}$ given by $\mathbf{r}_{1}$
and $\mathbf{r}_{2}$ are identical.
\end{thm}
\begin{IEEEproof}
By Lemma \ref{Generate.thm} write $\mathbf{Y}_{1}=\mathbf{F}_{1}^{-1}(\mathbf{U}_{1})$,
$\mathbf{Y}_{2}=\mathbf{F}_{2}^{-1}(\mathbf{U}_{2})$. Let $u$ be
the $k$-dimensional uniform pdf, i.e, $u(\mathbf{x})=1$ for $\mathbf{x}\in[0,1]^{k}$
and 0 otherwise, and let $\mathbf{Y}_{i}=\mathbf{s}_{i}^{-1}(\mathbf{t};\boldsymbol{\theta})$
denote the solution of $\mathbf{t}=\mathbf{s}_{i}(\mathbf{Y}_{i};\boldsymbol{\theta})$
with respect to $\mathbf{Y}_{i}$, which is a well-defined due to
assumption \ref{assumps.enu}. We can then write the distribution
of $\mathbf{t}$ in two ways as follows (\cite{GrimmettBook}), due
to the differentiability assumptions
\begin{align*}
f(\mathbf{t};\boldsymbol{\theta}) & =u(\mathbf{F}_{1}(\mathbf{s}_{1}^{-1}(\mathbf{t};\boldsymbol{\theta}))\left|\frac{\partial\mathbf{F}_{1}(\mathbf{s}_{1}^{-1}(\mathbf{t};\boldsymbol{\theta})}{\partial\mathbf{t}}\right|\\
 & =u(\mathbf{F}_{2}(\mathbf{s}_{2}^{-1}(\mathbf{t};\boldsymbol{\theta}))\left|\frac{\partial\mathbf{F}_{2}(\mathbf{s}_{2}^{-1}(\mathbf{t};\boldsymbol{\theta})}{\partial\mathbf{t}}\right|
\end{align*}
Due to assumption \ref{connected.enu} we can then that conclude $\frac{\partial\mathbf{F}_{1}(\mathbf{s}_{1}^{-1}(\mathbf{t};\boldsymbol{\theta})}{\partial\mathbf{t}}=\frac{\partial\mathbf{F}_{2}(\mathbf{s}_{2}^{-1}(\mathbf{t};\boldsymbol{\theta})}{\partial\mathbf{t}}$,
or
\[
\mathbf{F}_{1}(\mathbf{s}_{1}^{-1}(\mathbf{t};\boldsymbol{\theta})=\mathbf{F}_{2}(\mathbf{s}_{2}^{-1}(\mathbf{t};\boldsymbol{\theta}))+\mathbf{k}(\boldsymbol{\theta})
\]
But both $\mathbf{F}_{1}$ and $\mathbf{F}_{2}$ have range $[0,1]^{k}$,
and it follows that $\mathbf{k}(\boldsymbol{\theta})=\mathbf{0}$.
Therefore
\[
\mathbf{t}=\mathbf{s}_{1}(\mathbf{F}_{1}^{-1}(\mathbf{U});\boldsymbol{\theta})=\mathbf{s}_{2}(\mathbf{F}_{2}^{-1}(\mathbf{U});\boldsymbol{\theta})
\]
if we then solve either for $\boldsymbol{\theta}$ as a function of
$\mathbf{U}$ (for fixed $\mathbf{t}$), we therefore get exactly
the same result, and therefore the same distribution.
\end{IEEEproof}
The assumptions of Theorem \ref{thm:Uniqueness} are very restrictive,
but we believe they are far from necessary. In \cite{Sabeti2017ISITpredictive}
we proved uniqueness in the one-dimensional case under much weaker
assumptions (e.g., no differentiability assumptions), but that proof
is not easy to generalize to higher dimensions. 
\begin{cor}
\label{Equivalent.thm}Let $\mathbf{t}_{1}(x^{n})$ and $\mathbf{t}_{2}(x^{n})$
be \emph{equivalent }sufficient statistic for $\boldsymbol{\theta}$,
and assume the equivalence map is a diffeomorphism. Then the distribution
on $\boldsymbol{\theta}$ given by the sufficient statistic approach
is the same for $\mathbf{t}_{1}$ and $\mathbf{t}_{2}$.
\end{cor}
\begin{IEEEproof}
We have $\mathbf{t}_{1}=\mathbf{s}_{1}(\mathbf{Y}_{1},\boldsymbol{\theta})$
and $\mathbf{t}_{2}=\mathbf{s}_{2}(\mathbf{Y}_{2},\boldsymbol{\theta})$.
By assumption, there exists a one-to-one map $a$ so that $\mathbf{t}_{1}=a(\mathbf{t}_{2})$,
thus $\mathbf{t}_{1}=a(\mathbf{s}_{2}(\mathbf{Y}_{2},\boldsymbol{\theta}))$.
Since the distribution of $\boldsymbol{\theta}$ is independent of
how the problem is stated, $\mathbf{t}_{1}$ and $\mathbf{t}_{2}$
gives the same distribution on $\boldsymbol{\theta}$.
\end{IEEEproof}
We will compare the methods for a simple model. Assume our model is
$\mathcal{N}(0,\sigma^{2})$ with $\sigma$ unknown. The likelihood
function is $f(x^{n}|\sigma^{2})=\frac{1}{(2\pi\sigma^{2})^{n/2}}\exp\left(-\frac{1}{2\sigma^{2}}\sum_{i=1}^{n}x_{i}^{2}\right)$.
For $n=1$ we have$\int_{0}^{\infty}f(x^{n}|\sigma^{2})d\sigma^{2}=\infty$,
but for $n\geq2$ 
\begin{align*}
C\left(x^{n}\right) & =\int f(x^{n}|\sigma^{2})d\sigma^{2}=\frac{1}{\pi^{\frac{n}{2}}2}\frac{\Gamma\left(\frac{n-2}{2}\right)}{\left[n\widehat{\sigma^{2}}_{n}\right]^{\frac{n-2}{2}}}
\end{align*}
then 
\begin{align*}
f_{\text{nlm}}(x_{n+1}|x^{n}) & =\frac{\Gamma\left(\frac{n-1}{2}\right)}{\sqrt{\pi}\Gamma\left(\frac{n-2}{2}\right)}\frac{\left[n\widehat{\sigma^{2}}_{n}\right]^{\frac{n-2}{2}}}{\left[\left(n+1\right)\widehat{\sigma^{2}}_{n+1}\right]^{\frac{n-1}{2}}}
\end{align*}
where $\widehat{\sigma^{2}}_{n}=\frac{1}{n}\sum_{i=1}^{n}x_{i}^{2}$.
Thus, for coding, the two first samples would be encoded with the
default distribution, and after that the above distribution is used.
For the SSM, we note that $\widehat{\sigma^{2}}_{n}$ is a sufficient
statistic for $\sigma^{2}$ and that $z=\frac{n}{\sigma^{2}}\widehat{\sigma^{2}}_{n}\sim\chi_{(n)}^{2}$,
i.e., $\widehat{\sigma^{2}}_{n}=s(z,\sigma^{2})=\frac{\sigma^{2}}{n}z$,
which we can be solved as $\sigma^{2}=r(z,\widehat{\sigma^{2}}_{n})=\frac{n}{z}\widehat{\sigma^{2}}_{n}$,
in the notation of (\ref{tf.eq}-\ref{trY.eq}). This is a transformation
of the $\chi_{(n)}^{2}$ distribution which can be easily found as
\cite{GrimmettBook}
\begin{align*}
f_{\mathbf{x^{n}}}(\sigma^{2}) & =\frac{\left[n\widehat{\sigma^{2}}_{n}\right]^{\frac{n}{2}}}{2^{\frac{n}{2}}\Gamma\left(\frac{n}{2}\right)\left(\sigma^{2}\right)^{\frac{n+2}{2}}}\exp\left\{ -\frac{n}{2\sigma^{2}}\widehat{\sigma^{2}}_{n}\right\} 
\end{align*}
now we have  
\begin{align}
f_{\text{ssm}}(x_{n+1}|x^{n}) & =\int f(x_{n+1}|\sigma^{2})f_{x^{n}}(\sigma^{2})d\sigma^{2}\nonumber \\
 & =\frac{\Gamma\left(\frac{n+1}{2}\right)}{\sqrt{\pi}\Gamma\left(\frac{n}{2}\right)}\frac{\left[n\widehat{\sigma^{2}}_{n}\right]^{\frac{n}{2}}}{\left[\left(n+1\right)\widehat{\sigma^{2}}_{n+1}\right]^{\frac{n+1}{2}}}\label{eq:FssGauss}
\end{align}

\begin{figure}[htb]
\vspace{-0.15in}
 \hspace{-0.4in}\includegraphics[scale=0.7]{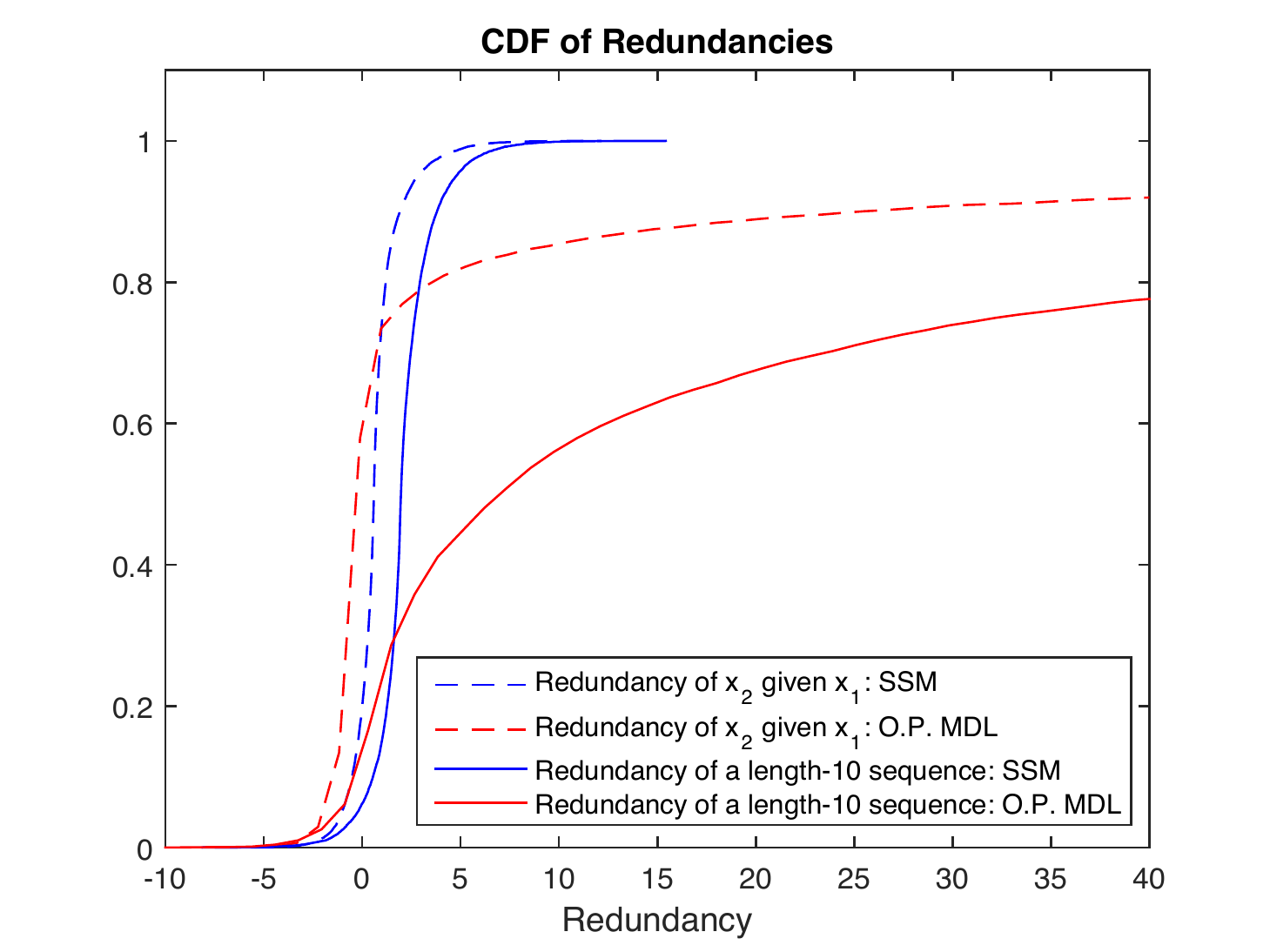} \caption{Redundancy comparison between ordinary predictive MDL (O.P. MDL) and
our proposed sufficient statistic method for $\mu=0$ and $\sigma^{2}=4$.}
\label{fig:Redundancy} \vspace{-0.2in}
 
\end{figure}

For comparison, the ordinary predictive MDL is
\begin{align}
f(x_{n+1}|x^{n}) & =\frac{1}{\sqrt{2\pi\widehat{\sigma^{2}}_{n}}}\exp\left(-\frac{1}{2\widehat{\sigma^{2}}_{n}}x_{n+1}^{2}\right)\label{PMDLGauss.eq}
\end{align}
which is of a completely different form. To understand the difference,
consider the codelength for $x_{2}$
\begin{align*}
L(x_{2}) & =\log\left(\frac{x_{1}^{2}+x_{2}^{2}}{|x_{1}|}\right)+\log\left(\frac{\sqrt{\pi}\Gamma(\frac{1}{2})}{\Gamma(1)}\right) & \text{SSM}\\
L(x_{2}) & =\frac{1}{2}\log\left(2\pi x_{1}^{2}\right)+\frac{x_{2}^{2}}{x_{1}^{2}} & \text{predictive MDL}
\end{align*}
At can be seen that if $x_{1}$ is small and $x_{2}$ is large, the
codelength for $x_{2}$ is going to be large. But in the sufficient
statistic method this is strongly attenuated due to the log in front
of the ratio. Fig. \ref{fig:Redundancy} shows this quantitatively
in the redundancy sense (difference between the codelength using true
and estimated distributions). As can be seen, the CDF of the ordinary
predictive MDL redundancy has a long tail, and this is taken care
of by SSM.

\section{\label{sec:Scalar}Scalar Signal Processing Methods}

In the following we will derive MDL for various scalar signal processing
methods. We can take inspiration from signal processing methods generally
used for source coding, such as linear prediction and wavelets; however,
the methods have to be modified for MDL, as we use lossless coding,
not lossy coding. As often in signal processing, the models are a
(deterministic) signal in Gaussian noise. In previous paper we have
also considered non-Gaussian models \cite{SabetiHost16Allerton}.
All proofs are in Appendices.

\subsection{\label{subsec:Iid-Gaussian-Case}Iid Gaussian Case }

\label{iidGaus.sec}A natural extension of the examples considered
in Section \ref{subsec:Method-of-Sufficient} is $x_{n}\sim\mathcal{N}(\mu,\sigma^{2})$
with both $\mu$ and $\sigma^{2}$ unknown. Define $\hat{\mu}_{n}=\frac{1}{n}\sum_{i=1}^{n}x_{i}$
and $S_{n}^{2}=\frac{1}{n-1}\sum_{i=1}^{n}\left(x_{i}-\hat{\mu}_{n}\right)^{2}$.
Then the sufficient statistic method is
\begin{align}
f(x_{n+1}|x^{n}) & =\sqrt{\frac{n}{\pi\left(n+1\right)}}\frac{\Gamma\left(\frac{n}{2}\right)}{\Gamma\left(\frac{n-1}{2}\right)}\nonumber \\
 & \times\frac{\left[\left(n-1\right)S_{n}^{2}\right]^{\frac{n-1}{2}}}{\left[nS_{n+1}^{2}\right]^{\frac{n}{2}}}\label{SSMmv.eq}
\end{align}
This is a special case of the vector Gaussian model considered later,
so we will not provide a proof. 

\subsubsection{Linear Transformations\label{lineartransform.sec}}

The iid Gaussian case is a fundamental building block for other MDL
methods. The idea is to find a linear transformation so that we can
model the result as iid, and then use the iid Gaussian MDL. For example,
in the vector case, suppose $\mathbf{x}_{n}\sim N(\boldsymbol{\mu},\boldsymbol{\Sigma})$
is (temporally) iid, and let $\mathbf{y}_{n}=\mathbf{A}\mathbf{x}_{n}\sim N(\mathbf{A}\boldsymbol{\mu},\mathbf{A}\boldsymbol{\Sigma}\mathbf{A}^{T})$.
If we then \emph{assume} that $\mathbf{A}\boldsymbol{\Sigma}\mathbf{A}^{T}$
is diagonal, we can use the iid Gaussian MDL on each component. Similarly,
in the scalar case, we can use a filter instead of a matrix. Because
of (\ref{scaling.eq}) we need to require $\mathbf{A}$ to be orthonormal:
for any input we then have $\mathbf{y}_{n}^{T}\mathbf{y}_{n}=\mathbf{x}_{n}^{T}\mathbf{A}^{T}\mathbf{A}\mathbf{x}_{n}=\mathbf{x}_{n}^{T}\mathbf{x}_{n}$,
and in particular $E[\mathbf{y}_{n}^{T}\mathbf{y}_{n}]=E[\mathbf{x}_{n}^{T}\mathbf{x}_{n}]$
independent of the actual $\boldsymbol{\Sigma}$. We will see this
approach in several cases in the following. 

\subsection{\label{subsec:LP}Linear Prediction}

Linear prediction is a fundamental to random processes. Write
\begin{align*}
\hat{x}_{n+1|x^{n}} & =\overset{\infty}{\underset{k=0}{\sum}}w_{k}x_{n-k}\\
e_{n+1} & =x_{n+1}-\hat{x}_{n+1|x^{n}}
\end{align*}
Then for most stationary random processes the resulting random process
$\{e_{n}\}$ is uncorrelated, and hence in the Gaussian case, iid,
by the Wold decomposition \cite{GrimmettBook}. It is therefore a
widely used method for source coding, e.g., \cite{GhidoTabus13}.
In practical coding, a finite prediction order $M$ is used, 

\begin{align*}
\hat{x}_{n+1|x^{n}} & =\overset{M}{\underset{k=1}{\sum}}w_{k}x_{n-k+1},\quad n\geq M
\end{align*}

Denote by $\tau$ the power of $\{e_{n}\}$. Consider the simplest
case with $M=1$: there are two unknown parameters $(w_{1},\tau)$.
However, the minimal sufficient statistic has dimension three \cite{Forchini00}:
$\left(\sum_{k=1}^{n}x_{k}^{2},\sum_{k=1}^{n-1}x_{k}^{2},\sum_{k=2}^{n}x_{k}x_{k-1}\right)$.
Therefore, we cannot use SSM; and even if we could, the distribution
of the sufficient statistic is not known in closed form \cite{Forchini00}.
We therefore turn to the NLM.

We assume that $e_{n+1}=x_{n+1}-\hat{x}_{n+1|x^{n}}$ is iid normally
distributed with zero mean and variance $\tau$, 

\begin{align}
f(x^{n}|\tau,\mathbf{w}) & =\frac{1}{(2\pi\tau)^{(n-M)/2}}\nonumber \\
 & \times\exp\left(-\frac{1}{2\tau}\sum_{i=M+1}^{n}\left[x_{i}-\sum_{k=1}^{M}w_{k}x_{i-k}\right]^{2}\right)\label{fxnLP.eq}
\end{align}
Define

\begin{align*}
\hat{r}_{(n)}(k) & =\sum_{i=M+1}^{n}x_{i}x_{i-k}
\end{align*}
Then a simple calculation shows that

\begin{align*}
\sum_{i=M+1}^{n}e_{i}^{2} & =\hat{r}_{(n)}(0)-2\mathbf{w}^{T}\mathbf{p}_{(n)}+\mathbf{w}^{T}R_{(n)}^{(M)}\mathbf{w}
\end{align*}
where $\mathbf{w}^{T}=[w_{1}\;w_{2}\;\cdots\;w_{M}]$, $\mathbf{p}_{(n)}^{T}=[\hat{r}_{(n)}(1)\;\hat{r}_{(n)}(2)\;\cdots\;\hat{r}_{(n)}(M)]$,
\begin{align}
R_{(n)}^{(M)} & =\sum_{i=M+1}^{n}\mathbf{x}_{i-M}^{i-1}\left(\mathbf{x}_{i-M}^{i-1}\right)^{T}\label{RLP.eq}
\end{align}
and $x_{i-M}^{i-1}=[x_{i-1},x_{i-2},\ldots,x_{i-M}]$. Thus 

\begin{align*}
f(x^{n}|\tau,\mathbf{w}) & =\frac{1}{(2\pi\tau)^{(n-M)/2}}\\
 & \times\exp\left(-\frac{1}{2\tau}\left[\hat{r}_{(n)}(0)-2\mathbf{w}^{T}\mathbf{p}_{(n)}+\mathbf{w}^{T}R_{(n)}^{(M)}\mathbf{w}\right]\right)
\end{align*}
giving (see Appendix \ref{Appendix: LinearPrediction})
\begin{align*}
C(x^{n}) & =\frac{1}{2\left(\pi\right)^{\frac{n-2M}{2}}\sqrt{\det\left(R_{(n)}\right)}}\frac{\Gamma\left(\frac{n-2M-2}{2}\right)}{\left(\hat{\tau}_{(n)}^{(M)}\right)^{\frac{n-2M-2}{2}}}
\end{align*}
and

\begin{align}
f_{M}(x_{n+1}|x^{n}) & =\sqrt{\frac{\det\left(R_{(n)}^{(M)}\right)}{\det\left(R_{(n+1)}^{(M)}\right)}}\frac{\Gamma\left(\frac{n-2M-1}{2}\right)}{\Gamma\left(\frac{n-2M-2}{2}\right)}\nonumber \\
 & \times\frac{1}{\sqrt{\pi}}\frac{\left(\hat{\tau}_{(n)}^{(M)}\right)^{\frac{n-2M-2}{2}}}{\left(\hat{\tau}_{(n+1)}^{(M)}\right)^{\frac{n-2M-1}{2}}}\label{NLMLP.eq}
\end{align}
with $\hat{\tau}_{(n)}^{(M)}=\hat{r}_{(n)}(0)-\mathbf{\mathbf{p}}_{(n)}^{T}R_{(n)}^{-1}\mathbf{p}_{(n)}$.

The equation (\ref{NLMLP.eq}) is defined for $n\geq2M+2$: the vector
$x_{i-M}^{i-1}$ is defined for $i\geq M+1$, and $R_{(n)}^{(M)}$
defined by (\ref{RLP.eq}) becomes full rank when the sum contains
$M$ terms. Before the order $M$ linear predictor becomes defined,
the data needs to be encoded with other methods. Since in atypicality
we are not seeking to determine the model of data, just if a different
model than the typical is better, we encode data with lower order
linear predictors until the order $M$ linear predictor becomes defined.
So, the first sample is encoded with the default pdf. The second and
third samples are encoded with the iid unknown variance coder (\ref{eq:FssGauss})\footnote{There is no issue in encoding some samples with SSM and others with
NLM.}. Then the order 1 linear predictor takes over, and so on. 

\subsection{\label{Filterbank.sec}Filterbanks and Wavelets}

A popular approach to source coding is subband coding and waveletts
\cite{mallat2008wavelet,VetterliWaveletsBook,VetterliHerley92}. The
basic idea is to divide the signal into (perhaps overlapping) spectral
subbands and then allocate different bitrates to each subband; the
bitrate can be dependent on the power in the subband and auditory
properties of the ear in for example audio coding. In MDL we need
to do lossless coding, so this approach cannot be directly applied,
but we can still use subband coding as explained in the following.

As we are doing lossless coding, we will only consider perfect reconstruction
filterbanks \cite{mitra2006DSP,mallat2008wavelet}. Furthermore, in
light of Section \ref{lineartransform.sec} we also consider only
(normalized) orthogonal filterbanks \cite{mallat2008wavelet,VetterliHerley92}.

The basic idea is that we split the signal into a variable number
of subbands by putting the signal through the filterbank and downsampling.
Then the output of each downsampled filter is coded with the iid Gaussian
coder of Section \ref{iidGaus.sec} with an unknown mean and variance
specific to each subband. In order to understand how this works, consider
a filterbank with two subbands. Assume that the signal is stationary
zero mean Gaussian with power $\sigma^{2}$, and let the power at
the output of subband 1 be $\sigma_{1}^{2}$ and of subband 2 be $\sigma_{2}^{2}$.
Because the filterbank is orthogonal, we have $\sigma^{2}=\frac{1}{2}\left(\sigma_{1}^{2}+\sigma_{2}^{2}\right)$.
For analysis purposes, using (\ref{RealMDL.eq}) it is straightforward
to see that we get the approximate codelengths
\begin{align*}
L_{\text{direct}} & =\frac{l}{2}\log\left(\sigma^{2}\right)+\frac{l}{2}\left(\log2\pi+\log e\right)+\frac{1}{2}\log l\\
L_{\text{filterbank}} & =\frac{l}{4}\log\left(\sigma_{1}^{2}\right)+\frac{l}{4}\left(\log2\pi+\log e\right)+\frac{1}{2}\log l\\
 & +\frac{l}{4}\log\left(\sigma_{1}^{2}\right)+\frac{l}{4}\left(\log2\pi+\log e\right)+\frac{1}{2}\log l\\
 & =\frac{l}{2}\log\left(\sqrt{\sigma_{1}^{2}\sigma_{2}^{2}}\right)+\frac{l}{2}\left(\log2\pi+\log e\right)+\log l
\end{align*}
Since $\sqrt{\sigma_{1}^{2}\sigma_{2}^{2}}\leq\sigma^{2}$ (with equality
only if $\sigma_{1}^{2}=\sigma_{2}^{2}$), the subband coder will
result in shorter codelength for sufficiently large $l$ if the signal
is non-white.

The above analysis is a stationary analysis for long sequences. However,
when considering shorter sequences, we also need to consider the transient.
The main issue is that output power will deviate from the stationary
value during the transient, and this will affect the estimated power
$\widehat{\sigma_{n}^{2}}$ used in the sequential MDL. The solution
is to transmit to the receiver the \emph{input} to the filterbank
during the transient, and only use the output of the filterbank once
the filters have been filled up. It is easy to see that the system
is still perfect reconstruction: Using the received input to the filterbank,
the receiver puts this through the \emph{analysis} filterbank. It
now has the total sequence produced by the analysis filterbank, and
it can then put that through the reconstruction filterhank. When using
multilevel filterbanks, this has to be done at each level.

We assume the decoder knows which filters are used and the maximum
depth $D$ used. In principle the encoder could now search over all
trees of level at most $D$. The issue is that there are an astonishing
large number of such trees; for example for $D=4$ there are 676 such
trees. Instead of choosing the best, we can use the idea of the CTW
\cite{WillemsAl95,WillemsAl97,HostSabetiWalton15} and weigh in each
node: Suppose after passing a signal $x^{n}$ of an internal node
$S$ through low-pass and high-pass filters and downsampler, $x_{L}^{n/2}$
and $x_{H}^{n/2}$ are produced in the children nodes of $S$. The
weighted probability of $x^{n}$ in the internal node $S$ will be
\begin{align*}
f_{w}\left(x^{n}\right) & =\frac{1}{2}f\left(x^{n}\right)+\frac{1}{2}f_{w}\left(x_{L}^{n/2}\right)f_{w}\left(x_{H}^{n/2}\right)
\end{align*}
which is a good coding distribution for both a memoryless source and
a source with memory \cite{WillemsAl95,WillemsAl97}.

\section{Vector Case}

We now assume that a vector sequence $\mathbf{x}^{n}$, $\mathbf{x}_{i}\in\mathbb{R}^{M}$
is observed. The vector case allows for a more rich set of model and
more interesting data discovery than the scalar case, for example
atypical correlation between multiple sensors. It can also be applied
to images \cite{SabetiHost16BD}, and to scalar data by dividing into
blocks. That is in particular useful for the DFT, Section \ref{subsec:DFT}.

A specific concern is initialization. Applying sequential coding verbatim
to the vector case means that the first vector $\mathbf{x}_{1}$ need
to be encoded with the default coder, but this means the default coder
influences codelength too much. Instead we suggest to encode the first
vector as a scalar signal using the scalar Gaussian coder (unknown
variance$\to$unknown mean/variance). That way only the first component
of the first vector needs to be encoded with the default coder.

\subsection{Vector Gaussian Case with Unknown $\boldsymbol{\mu}$}

First assume $\boldsymbol{\mu}$ is unknown but $\Sigma$ is given.
We define $\mathrm{etr}\left(\cdots\right)=\exp\left(\mathrm{trace}\left(\cdots\right)\right)$
and we have

\begin{align*}
f\left(\mathbf{x}^{n}|\boldsymbol{\mu}\right) & =\frac{1}{\sqrt{\left(2\pi\right)^{kn}\det\left(\Sigma\right)^{n}}}\\
 & \times\exp\left\{ -\frac{1}{2}\sum_{i=1}^{n}\left(\mathbf{x}_{i}-\boldsymbol{\mu}\right)^{T}\Sigma^{-1}\left(\mathbf{x}_{i}-\boldsymbol{\mu}\right)\right\} 
\end{align*}
We first consider the NLM. By defining $\hat{\boldsymbol{\mu}}_{n}=\frac{1}{n}\sum_{i=1}^{n}\mathbf{x}_{i}$
and $\hat{\Sigma}_{n}=\sum_{i=1}^{n}\mathbf{x}_{i}\mathbf{x}_{i}$
(note that $\hat{\Sigma}_{n}$ is not the estimate of $\Sigma$) we
have
\begin{align*}
C\left(\mathbf{x}^{n}\right) & =\int f\left(\mathbf{x}^{n}|\boldsymbol{\mu}\right)d\boldsymbol{\mu}\\
 & =\frac{1}{\sqrt{\left(2\pi\right)^{kn}\det\left(\Sigma\right)^{n}}}\exp\left\{ -\frac{1}{2}\sum_{i=1}^{n}\mathbf{x}_{i}\Sigma^{-1}\mathbf{x}_{i}\right\} \\
 & \times\int\exp\left\{ -\frac{n}{2}\boldsymbol{\mu}^{T}\Sigma^{-1}\boldsymbol{\mu}+n\hat{\boldsymbol{\mu}}_{n}^{T}\Sigma^{-1}\boldsymbol{\mu}\right\} d\boldsymbol{\mu}\\
 & =C\exp\left\{ -\frac{1}{2}\sum_{i=1}^{n}\left(\mathbf{x}_{i}\Sigma^{-1}\mathbf{x}_{i}-\hat{\boldsymbol{\mu}}_{n}^{T}\Sigma^{-1}\hat{\boldsymbol{\mu}}_{n}\right)\right\} \\
 & =C\mathrm{etr}\left\{ -\frac{1}{2}\left(\hat{\Sigma}_{n}-n\hat{\boldsymbol{\mu}}_{n}\hat{\boldsymbol{\mu}}_{n}^{T}\right)\Sigma^{-1}\right\} 
\end{align*}
where $C=\frac{1}{\sqrt{\left(2\pi\right)^{k\left(n-1\right)}n^{k}\det\left(\Sigma\right)^{n-1}}}$
hence we can write
\begin{align}
\lefteqn{f\left(\mathbf{x}_{n+1}|\mathbf{x}^{n}\right)=\frac{C\left(\mathbf{x}^{n+1}\right)}{C\left(\mathbf{x}^{n}\right)}}\nonumber \\
 & =\sqrt{\left(\frac{n}{n+1}\right)^{k}}\frac{1}{\sqrt{\left(2\pi\right)^{k}\det\left(\Sigma\right)}}\nonumber \\
 & \times\frac{\mathrm{etr}\left\{ -\frac{1}{2}\left(\hat{\Sigma}_{n+1}-\left(n+1\right)\hat{\boldsymbol{\mu}}_{n+1}\hat{\boldsymbol{\mu}}_{n+1}^{T}\right)\Sigma^{-1}\right\} }{\mathrm{etr}\left\{ -\frac{1}{2}\left(\hat{\Sigma}_{n}-n\hat{\boldsymbol{\mu}}_{n}\hat{\boldsymbol{\mu}}_{n}^{T}\right)\Sigma^{-1}\right\} }\label{VGmNL.eq}
\end{align}
It turns out that in this case, the SSM gives the same result.

\subsection{\label{subsec:VecGausCov}Vector Gaussian Case with Unknown $\Sigma$}

Assume $\mathbf{x}_{n}\sim\mathcal{N}\left(\mathbf{0},\Sigma\right)$
where the covariance matrix is unknown
\begin{align*}
f\left(\mathbf{x}^{n}|\Sigma\right) & =\frac{1}{\sqrt{\left(2\pi\right)^{kn}\det\left(\Sigma\right)^{n}}}\mathrm{etr}\left\{ -\frac{1}{2}\hat{\Sigma}_{n}\Sigma^{-1}\right\} 
\end{align*}
where $\hat{\boldsymbol{\Sigma}}_{n}=\sum_{i=1}^{n}\mathbf{x}_{i}\mathbf{x}_{i}^{T}$.

In order to find the MDL using SSM, notice that we can write
\[
\mathbf{x}_{n}=\mathbf{S}\mathbf{z}_{n},\quad\mathbf{z}_{n}\sim\mathcal{N}(0,\mathbf{I})
\]
where $\mathbf{S}=\boldsymbol{\Sigma}^{\frac{1}{2}}$, that is $\mathbf{S}$
is \emph{some} matrix that satisfies $\mathbf{S}\mathbf{S}^{T}=\boldsymbol{\Sigma}$.
A sufficient statistic for $\boldsymbol{\Sigma}$ is 
\[
\hat{\boldsymbol{\Sigma}}_{n}=\sum_{i=1}^{n}\mathbf{x}_{i}\mathbf{x}_{i}^{T}=\mathbf{S}\sum_{i=1}^{n}\mathbf{z}_{i}\mathbf{z}_{i}^{T}\mathbf{S}^{T}\stackrel{\text{def}}{=}\mathbf{S}\mathbf{U}\mathbf{S}^{T}
\]
Let $\hat{\mathbf{S}}_{n}=\hat{\boldsymbol{\Sigma}}_{n}^{\frac{1}{2}}=\mathbf{S}\mathbf{U}^{\frac{1}{2}}$.
Then we can solve $\mathbf{S}=\hat{\mathbf{S}}_{n}\mathbf{U}^{-\frac{1}{2}}$
and $\boldsymbol{\Sigma}=\mathbf{\hat{S}}_{n}\mathbf{U}^{-1}\mathbf{\hat{S}}_{n}^{T}$.
Since $\mathbf{U}^{-1}$ has Inverse-Wishart distribution $\mathbf{U}^{-1}\sim\mathcal{W}_{M}^{-1}\left(I,n\right)$,
one can write $\Sigma\sim\mathcal{W}_{M}^{-1}\left(\hat{\Sigma}_{n},n\right)$.
Using this distribution we calculate in Appendix \ref{Appendix:VecGausVar}
that 
\begin{align}
f\left(\mathbf{x}_{n+1}|\mathbf{x}^{n}\right) & =\frac{1}{\pi^{\frac{M}{2}}}\frac{\det\left(\hat{\Sigma}_{n}\right)^{\frac{n}{2}}}{\det\left(\hat{\Sigma}_{n+1}\right)^{\frac{n+1}{2}}}\frac{\Gamma_{M}\left(\frac{n+1}{2}\right)}{\Gamma_{M}\left(\frac{n}{2}\right)}\label{VGSSS.eq}
\end{align}
where $\Gamma_{M}$ is the multivariate gamma function \cite{muirhead2009MultivariateStat}.

On the other hand, using the normalized likelihood method we have
\begin{align*}
C\left(\mathbf{x}^{n}\right) & =\frac{\Gamma_{M}\left(\frac{n}{2}-\frac{M+1}{2}\right)}{2^{\frac{M\left(M+1\right)}{2}}\pi^{\frac{kn}{2}}\det\left(\hat{\Sigma}_{n}\right)^{\frac{n}{2}-\frac{M+1}{2}}}
\end{align*}
From which 
\begin{align}
f\left(\mathbf{x}_{n+1}|\mathbf{x}^{n}\right) & =\frac{C\left(\mathbf{x}^{n+1}\right)}{C\left(\mathbf{x}^{n}\right)}\nonumber \\
 & =\frac{1}{\pi^{\frac{k}{2}}}\frac{\det\left(\hat{\Sigma}_{n}\right)^{\frac{n}{2}-\frac{M+1}{2}}}{\det\left(\hat{\Sigma}_{n+1}\right)^{\frac{n}{2}-\frac{M}{2}}}\frac{\Gamma_{M}\left(\frac{n}{2}-\frac{M}{2}\right)}{\Gamma_{M}\left(\frac{n}{2}-\frac{M+1}{2}\right)}\label{VGSNL.eq}
\end{align}

\subsection{Vector Gaussian Case with Unknown $\boldsymbol{\mu}$ and $\Sigma$}

Assume $\mathbf{x}_{n}\sim\mathcal{N}\left(\boldsymbol{\mu},\Sigma\right)$
where both mean and covariance matrix are unknown

\begin{align*}
f\left(\mathbf{x}^{n}|\boldsymbol{\mu},\Sigma\right) & =\frac{1}{\sqrt{\left(2\pi\right)^{Mn}\det\left(\Sigma\right)^{n}}}\\
 & \times\exp\left\{ -\frac{1}{2}\sum_{i=1}^{n}\left(\mathbf{x}_{i}-\boldsymbol{\mu}\right)^{T}\Sigma^{-1}\left(\mathbf{x}_{i}-\boldsymbol{\mu}\right)\right\} 
\end{align*}
It is well-known \cite{ScharfBook} that sufficient statistics are
$\hat{\boldsymbol{\mu}}_{n}=\frac{1}{n}\sum_{i=1}^{n}\mathbf{x}_{i}$
and $\hat{\boldsymbol{\Sigma}}_{n}=\left(n-1\right)S_{n}=\sum_{i=1}^{n}\left(\mathbf{x}_{i}-\hat{\boldsymbol{\mu}}_{n}\right)\left(\mathbf{x}_{i}-\hat{\boldsymbol{\mu}}_{n}\right)^{T}$.
Let $\mathbf{S}$ be a square root of $\boldsymbol{\Sigma}$, i.e.,
$\mathbf{S}\mathbf{S}^{T}=\boldsymbol{\Sigma}$. We can then write
\begin{align*}
\hat{\boldsymbol{\mu}}_{n} & =\boldsymbol{\mu}+\frac{1}{\sqrt{n}}\mathbf{S}\mathbf{z}\\
\hat{\boldsymbol{\Sigma}}_{n} & =\mathbf{S}\mathbf{U}\mathbf{S}^{T}
\end{align*}
where $\mathbf{z}\sim\mathcal{N}\left(\mathbf{0},I\right)$ and $\mathbf{U}\sim\mathcal{W}_{M}\left(\mathbf{I},n-1\right)$,\textbf{
$\mathbf{z}$ }and $\mathbf{U}$ are independent, and $\mathcal{W}_{M}$
is the Wishart distribution. We solve the second equation with respect
to $\mathbf{S}$ as in Section \ref{subsec:VecGausCov} and the first
with respect to $\boldsymbol{\mu}$, to get
\begin{align*}
\boldsymbol{\Sigma} & =\hat{\mathbf{S}}_{n}\mathbf{U}^{-1}\hat{\mathbf{S}}_{n}^{T}\sim\mathcal{W}_{M}^{-1}\left(\hat{\boldsymbol{\Sigma}}_{n},n-1\right)\\
\boldsymbol{\mu} & =\hat{\boldsymbol{\mu}}_{n}-\frac{1}{\sqrt{n}}\mathbf{S}\mathbf{z}=\hat{\boldsymbol{\mu}}_{n}-\frac{1}{\sqrt{n}}\hat{\mathbf{S}}_{n}\mathbf{U}^{-\frac{1}{2}}\mathbf{z}\sim\mathcal{N}\left(\hat{\boldsymbol{\mu}}_{n},\frac{1}{n}\Sigma\right)
\end{align*}
where $\hat{\mathbf{S}}_{n}$ is a square root of $\hat{\boldsymbol{\Sigma}}_{n}$.
We can explicitly write the distributions as 
\begin{align*}
f_{\mathbf{x}^{n}}\left(\boldsymbol{\mu}|\Sigma\right) & =\sqrt{\frac{n^{M}}{\left(2\pi\right)^{M}\det\left(\Sigma\right)}}\exp\left\{ -\frac{n}{2}\left(\boldsymbol{\mu}-\hat{\boldsymbol{\mu}}_{n}\right)^{T}\Sigma^{-1}\left(\boldsymbol{\mu}-\hat{\boldsymbol{\mu}}_{n}\right)\right\} \\
f_{\mathbf{x}^{n}}\left(\Sigma\right) & =\frac{\det\left(\hat{\Sigma}_{n}\right)^{\frac{n-1}{2}}}{2^{\frac{M\left(n-1\right)}{2}}\Gamma_{M}\left(\frac{n-1}{2}\right)}\det\left(\Sigma\right)^{-\frac{n+M}{2}}\mathrm{etr}\left\{ -\frac{1}{2}\hat{\Sigma}_{n}\Sigma^{-1}\right\} 
\end{align*}
Using these distributions, in Appendix \ref{Appendix:VecGausMeanVar}
we calculate
\begin{align*}
f\left(\mathbf{x}_{n+1}|\mathbf{x}^{n}\right) & =\frac{1}{\pi^{\frac{M}{2}}}\sqrt{\left(\frac{n}{n+1}\right)^{M}}\frac{\det\left(\hat{\Sigma}_{n}\right)^{\frac{n-1}{2}}}{\det\left(\hat{\Sigma}_{n+1}\right)^{\frac{n}{2}}}\frac{\Gamma_{M}\left(\frac{n}{2}\right)}{\Gamma_{M}\left(\frac{n-1}{2}\right)}
\end{align*}
and for NLM
\begin{align*}
f\left(\mathbf{x}_{n+1}|\mathbf{x}^{n}\right) & =\frac{1}{\pi^{\frac{M}{2}}}\sqrt{\left(\frac{n}{n+1}\right)^{M}}\frac{\det\left(\hat{\Sigma}_{n}\right)^{\frac{n-1}{2}-\frac{M+1}{2}}}{\det\left(\hat{\Sigma}_{n+1}\right)^{\frac{n}{2}-\frac{M+1}{2}}}\\
 & \times\frac{\Gamma_{M}\left(\frac{n-M-1}{2}\right)}{\Gamma_{M}\left(\frac{n-M-2}{2}\right)}
\end{align*}
These are very similar to the case of known mean, Section \ref{subsec:VecGausCov}.
We require one more sample before the distributions become well-defined,
and $\boldsymbol{\Sigma}_{n}$ is defined differently. 

\subsection{\label{subsec:DFT}Sparsity and DFT}

We can specify a general method as follows. Let $\boldsymbol{\Phi}$
is an orthonormal basis of $\mathbb{R}^{M}$ and write the signal
model as
\[
\mathbf{x}_{n}=\sum_{i=1}^{N}(A_{i}+s_{i,n})\boldsymbol{\phi}_{j(i)}+\mathbf{w}_{n}
\]
Here $N$ is the number of basis vectors used, and $j(i),i=1,\ldots,N$
their indices. The signal $s_{i,n}$ is iid $\mathcal{N}(0,\sigma_{i})$,
the noise $\mathbf{w}_{n}$ iid $\mathcal{N}(0,\sigma^{2}\mathbf{I})$,
and $A_{i},\sigma_{i}^{2},\sigma^{2}$ are unknown. If we let $\mathbf{y}_{n}=\boldsymbol{\Phi}^{T}\mathbf{x}_{n}$
and $J$ the indices of the signal components then
\begin{align*}
y_{j(i),n} & =A_{i}+s_{i,n}+w_{j(i),n}=A_{i}+\tilde{s}_{i,n},\quad j(i)\in J\\
y_{j,n} & =w_{j,n},\quad j\notin J
\end{align*}
Thus the $y_{j(i),n}$ can be encoded with the scalar Gaussian encoder
of Section \ref{iidGaus.sec}, while the $y_{j,n}$ can be encoded
with a vector Gaussian encoder for $\mathcal{N}(0,\sigma^{2}\mathbf{I}_{M-N})$
using the following equation that is achieved using the SSM 
\begin{align*}
f\left(\mathbf{w}_{n+1}|\mathbf{w}^{n}\right) & =\frac{1}{\pi^{\frac{\left(M-N\right)}{2}}}\frac{\Gamma\left(\frac{\left(M-N\right)\left(n+1\right)}{2}\right)}{\Gamma\left(\frac{\left(M-N\right)n}{2}\right)}\\
 & \times\frac{\left[n\widehat{\tau}_{n}\right]^{\frac{\left(M-N\right)n}{2}}}{\left[\left(n+1\right)\widehat{\tau}_{n+1}\right]^{\frac{\left(M-N\right)\left(n+1\right)}{2}}}
\end{align*}
where $\widehat{\tau}_{n}=\frac{1}{n}\sum_{i=1}^{n}\mathbf{w}_{i}^{T}\mathbf{w}_{i}$.
Now we need to choose which coefficients $j(i)$ to choose as signal
components and inform the decoder. The set $J$ can be communicated
to the decoder by sending a sequence of $0,1$ encoded with the universal
encoder of \cite[Section 13.2]{CoverBook} with $MH\left(\frac{N}{M}\right)+\frac{1}{2}\log M$
bits. The optimum set can in general only be found by trying all sets
$J$ and choosing the one with shortest codelength, which is infeasible.
A heuristic approach is to find the $N$ components with maximum power
when calculated\footnote{The decoder does not need to know how $J$ was chosen, only what $J$
is. It is therefore fine to use the power at the end of the block.} over the whole blocklength $l$. What still remains is how to choose
$N$. It seems computationally feasible to start with $N=1$ and then
increase $N$ by 1 until the codelength no longer decreases, since
most of the calculations for $N$ can be reused for $N+1$.

We can apply this in particular when $\boldsymbol{\Phi}$ is a DFT
matrix. In light of Section \ref{lineartransform.sec} we need to
use the normalized form of the DFT. The complications is that output
is complex, i.e., the $M$ real inputs result in $M$ complex outputs,
or $2M$ real outputs. Therefore, care has to be taken with the symmetry
properties of the output. Another option is to use DCT instead, which
is well-developed and commonly used for compression.

\section{\label{sec:Experimental-Results}Experimental Results}

As an example of application of atypicality, we will consider transient
detection \cite{ChenWillettAl98}. In transient detection, a sensor
records a signal that is pure noise most of the time, and the task
is to find the sections of the signal that are not noise. In our terminology,
the typical signal is noise, and the task is to find the atypical
parts. 

As data we used hydrophone recordings from a sensor in the Hawaiian
waters outside Oahu, the Station ALOHA Cabled Observatory (\textquotedblleft ACO\textquotedblright )
\cite{KaraSilver2014Ocean}. The data used for this paper were collected
(with sampling freuquency of 96 kHz which was then downsampled to
8 kHz) during a proof module phase of the project conducted between
February 2007 and October 2008. The data was pre-processed by differentiation
($y[n]=x[n]-x[n-1])$ to remove a non-informative mean component.

The principal goal of this two years of data is to locate whale vocalization.
Fin (22 meters, up to 80 tons) and sei (12-18 meters, up to 24.6 tons)
whales are known by means of visual and acoustic surveys to be present
in the Hawaiian Islands during winter and spring months, but migration
patterns in Hawaii are poorly understood \cite{KaraSilver2014Ocean}.

Ground truth has been established by manual detection, which is achieved
using visual inspection of spectrogram by a human operator. 24 hours
of manual detections for both the 20 Hz and the 20-35 Hz variable
calls were recorded for each the following dates (randomly chosen):
01 March 2007, 17 November 2007, 29 May 2008, 22 August 2008, 04 September
2008 and 09 February 2008 \cite{KaraSilver2014Ocean}. 

In order to analyze the performance of different detectors on such
a data, first the measures Precision and Recall are defined as below
\begin{align*}
\text{Recall} & =\frac{\text{number\,of\,correct\,detection}s}{\text{total\,number\,of\,manual\,detections}}\\
\text{Precision} & =\frac{\text{number\,of\,correct\,detections}}{\text{total\,number\,of\,algorithm\,detections}}
\end{align*}
where Recall measures the probability of correctly obtained vocalizations
over expected number of detections and Precision measures the probability
of correctly detected vocalizations obtained by the detector. The
Precision versus Recall curve show the detectors ability to obtain
vocalizations as well as the accuracy of these detections \cite{KaraSilver2014Ocean}.

In order to compare our atypicality method with alternative approaches
in transient detection, we compare its performance with Variable Threshold
Page (VTP) which outperforms other similar methods in detection of
non-trivial signals \cite{willett2005VTP}. 

For the atypicality approach, we need a typical and and an atypical
coder. The typical signal is pure noise, which, however, is not necessarily
white: it consists of background noise, wave motion, wind and rain.
We therefore used a linear predictive coder. The order of the linear
predictive coder was globally set to 10 as a compromise between performance
and computational speed. An order above 10 showed no significant decrease
in codelength, while increasing computation time. The prediction coefficients
were estimated for each 5-minute segment of data. It seems unreasonable
to expect the prediction coefficients to be globally constant due
to for example variations in weather, but over short length segments
they can be expected to be constant. Of course, a 5 minute segment
could contain atypical data and that would result in incorrect typical
prediction coefficients. However, for this particular data we know
(or assume) that atypical segments are of very short duration, and
therefore will affect the estimated coefficients very little. This
cannot be used for general data sets, only for data sets where there
is a prior knowledge (or assumption) that atypical data are rare and
short. Otherwise the typical coder should be trained on data known
to be typical as in \cite{HostSabetiWalton15} or by using unsupervised
atypicality \cite{SabetiHost17BD}.

For the atypical coder, we implemented all the scalar methods of section
\ref{sec:Scalar} in addition to the DFT, Section \ref{subsec:DFT},
with optimization over blocklength. Searching for atypical sequence
(in this case, whale vocalizations) was then performed in different
stages (more details of algorithm implementation can be found in \cite{Host15ISIT}).
Fig. \ref{fig:PreVsRecall} shows Precision vs Recall curve for both
atypicality and VTP. 
\begin{center}
\begin{figure}[tbh]
\begin{centering}
\includegraphics[scale=0.6]{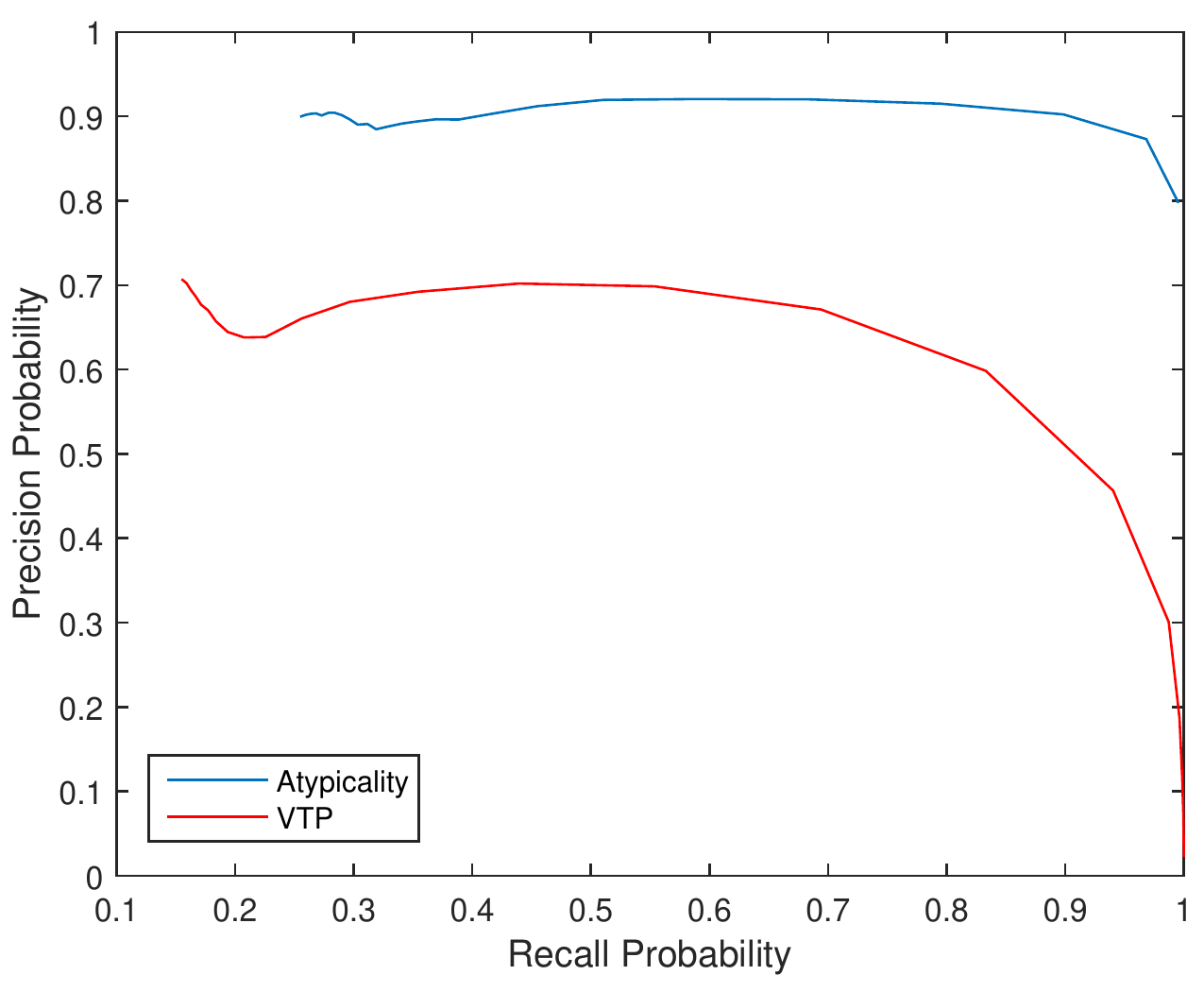}\vspace{-0.15in}
\par\end{centering}
\begin{centering}
\caption{\label{fig:PreVsRecall} Precision vs Recall probability for all six
days that manual detections are available}
\par\end{centering}
\centering{}\vspace{-0.1in}
\end{figure}
\par\end{center}

\section{Conclusion}

Atypicality is a method for finding rare, interesting snippets in
big data. It can be used for anomaly detection, data mining, transient
detection, and knowledge extraction among other things. The current
paper extended atypicality to real-valued data. It is important here
to notice that discrete-valued and real-valued atypicality is one
theory. Atypicality can therefore be used on data that are of mixed
type. One advantage of atypicality is that it directly applies to
sequences of variable length. Another advantage is that there is only
one parameter that regulates atypicality, the single threshold parameter
$\tau$, which has the concrete meaning of the logarithm of the frequency
of atypical sequences. This contrasts with other methods that have
multiple parameters.

Atypicality becomes really interesting in combination with machine
learning. First, atypicality can be used to find what is \emph{not}
learned in machine learning. Second, for many data sets, machine learning
is needed to find the typical coder. In the experiments in this paper,
we did not need machine learning because the typical data was pure
noise. But in many other types of data, e.g., ECG (electrocardiogram),
``normal'' data is highly complex, and the optimum coder has to
be learned with machine learning. This is a topic for future research.

\bibliographystyle{IEEEtran}
\bibliography{Coop06,Sensor,ahmref2,Coop03,BigData,Underwater,CDMA,combined,ECGandHRV,nosal}

\appendices{}

\section{\label{Appendix: LinearPrediction}Linear Prediction}

we showed
\begin{align*}
f(x^{n}|\tau,\mathbf{w}) & =\frac{1}{(2\pi\tau)^{(n-M)/2}}\\
 & \times\exp\left(-\frac{1}{2\tau}\left[\hat{r}_{(n)}(0)-2\mathbf{w}^{T}\mathbf{p}_{(n)}+\mathbf{w}^{T}R_{(n)}^{(M)}\mathbf{w}\right]\right)
\end{align*}
therefore using NLM we have
\begin{align*}
C(x^{n}) & =\int\int f(x^{n}|\tau,\mathbf{w})d\mathbf{w}d\tau\\
 & =A\int_{\tau}\tau^{-\frac{\left(n-M\right)}{2}}\exp\left\{ -\frac{\hat{r}_{(n)}(0)}{2\tau}\right\} e_{1}\left(\tau\right)d\tau
\end{align*}
where $A=\frac{1}{(2\pi)^{\left(n-M\right)/2}}$ and $e_{1}\left(\tau\right)=\int_{\mathbf{w}}\exp\left\{ -\frac{1}{2\tau}\left[\mathbf{w}^{T}R_{(n)}\mathbf{w-}2\mathbf{p}_{(n)}^{T}\mathbf{w}\right]\right\} d\mathbf{w}$.
Hence
\begin{align*}
C(\mathbf{x}^{n}) & =B\int_{\tau}\tau^{-\frac{n-2M}{2}}\exp\left\{ -\frac{1}{2\tau}\left[\hat{r}_{(n)}(0)-\mathbf{\mathbf{p}}_{(n)}^{T}R_{(n)}^{-1}\mathbf{p}_{(n)}\right]\right\} d\tau\\
 & =B\int_{\tau}\tau^{-\frac{n-2M}{2}}\exp\left\{ -\frac{1}{2\tau}\hat{\tau}_{(n)}^{(M)}\right\} d\tau\\
 & =\frac{1}{2\left(\pi\right)^{\frac{n-2M}{2}}\sqrt{\det\left(R_{(n)}\right)}}\frac{\Gamma\left(\frac{n-2M-2}{2}\right)}{\left(\hat{\tau}_{(n)}^{(M)}\right)^{\frac{n-2M-2}{2}}}
\end{align*}
where $B=\frac{1}{(2\pi)^{\left(n-2M\right)/2}\sqrt{\det\left(R_{(n)}\right)}}$.

\section{\label{Appendix:VecGausVar}Vector Gaussian Case: unknown $\Sigma$}

We showed that $\Sigma$ has Inverse-Wishart distribution $\Sigma\sim\mathcal{W}_{M}^{-1}\left(\hat{\Sigma}_{n},n\right)$
where $\hat{\Sigma}_{n}=\sum_{i=1}^{n}\mathbf{x}_{i}\mathbf{x}_{i}$
, hence

\begin{align*}
f_{\mathbf{x}^{n}}\left(\Sigma\right) & =\frac{\det\left(\hat{\Sigma}_{n}\right)^{\frac{n}{2}}}{2^{\frac{nM}{2}}\Gamma_{M}\left(\frac{n}{2}\right)}\det\left(\Sigma\right)^{-\frac{n+M+1}{2}}\mathrm{etr}\left\{ -\frac{1}{2}\hat{\Sigma}_{n}\Sigma^{-1}\right\} 
\end{align*}
and since
\begin{align*}
f\left(\mathbf{x}_{n+1}|\Sigma\right) & =\frac{1}{\sqrt{\left(2\pi\right)^{M}\det\left(\Sigma\right)}}\mathrm{etr}\left\{ -\frac{1}{2}\left(\hat{\Sigma}_{n+1}-\hat{\Sigma}_{n}\right)\Sigma^{-1}\right\} 
\end{align*}
therefore we have
\begin{align*}
f\left(\mathbf{x}_{n+1}|\mathbf{x}^{n}\right) & =\int_{\Sigma>0}f\left(\mathbf{x}_{n+1}|\Sigma\right)f_{\mathbf{x}^{n}}\left(\Sigma\right)d\Sigma\\
 & =C\int_{\Sigma>0}\det\left(\Sigma\right)^{-\frac{n+M+2}{2}}\mathrm{etr}\left\{ -\frac{1}{2}\hat{\Sigma}_{n+1}\Sigma^{-1}\right\} d\Sigma\\
 & \overset{(A)}{=}C\int_{Y>0}\det\left(Y\right)^{\frac{n}{2}-\frac{M}{2}}\mathrm{etr}\left\{ -\frac{1}{2}\hat{\Sigma}_{n+1}Y\right\} dY\\
 & =D\int_{V>0}\det\left(V\right)^{\frac{n}{2}-\frac{M}{2}}\mathrm{etr}\left\{ -V\right\} dV\\
 & \overset{(B)}{=}D\int_{V>0}\det\left(V\right)^{\frac{n+1}{2}-\frac{M+1}{2}}\mathrm{etr}\left\{ -V\right\} dV\\
 & =\frac{1}{\pi^{\frac{M}{2}}}\frac{\det\left(\hat{\Sigma}_{n}\right)^{\frac{n}{2}}}{\det\left(\hat{\Sigma}_{n+1}\right)^{\frac{n+1}{2}}}\frac{\Gamma_{M}\left(\frac{n+1}{2}\right)}{\Gamma_{M}\left(\frac{n}{2}\right)}
\end{align*}
where $C=\frac{\det\left(\hat{\Sigma}_{n}\right)^{\frac{n}{2}}}{2^{\frac{M\left(n+1\right)}{2}}\Gamma_{M}\left(\frac{n}{2}\right)\pi^{\frac{M}{2}}}$
and $D=\frac{\det\left(\hat{\Sigma}_{n}\right)^{\frac{n}{2}}}{\det\left(\hat{\Sigma}_{n+1}\right)^{\frac{n+1}{2}}}\frac{1}{\Gamma_{M}\left(\frac{n}{2}\right)\pi^{\frac{M}{2}}}$,
and in equations (A) and (B) we changed the variable $\Sigma=Y^{-1}$
and $Y=2\hat{\Sigma}_{n}^{-\frac{1}{2}}V\hat{\Sigma}_{n}^{-\frac{1}{2}}$
respectively and $\Gamma_{m}\left(a\right)=\int_{V>0}\det\left(V\right)^{a-\frac{\left(m+1\right)}{2}}\mathrm{etr}\left\{ -V\right\} dV$
is the multivariate Gamma function.

\section{\label{Appendix:VecGausMeanVar}Vector Gaussian Case: unknown $\boldsymbol{\mu}$
and $\Sigma$}

We showed that $\boldsymbol{\mu}\sim\mathcal{N}\left(\hat{\boldsymbol{\mu}}_{n},\frac{1}{n}\Sigma\right)$
and $\Sigma\sim\mathcal{W}_{M}^{-1}\left(\hat{\Sigma}_{n},n-1\right)$
where $\hat{\boldsymbol{\mu}}_{n}=\frac{1}{n}\sum_{i=1}^{n}\mathbf{x}_{i}$
and $\hat{\boldsymbol{\Sigma}}_{n}=\sum_{i=1}^{n}\left(\mathbf{x}_{i}-\hat{\boldsymbol{\mu}}_{n}\right)\left(\mathbf{x}_{i}-\hat{\boldsymbol{\mu}}_{n}\right)^{T}$.
Now using Bayes we can write the joint pdf as $f_{\mathbf{x}^{n}}\left(\boldsymbol{\mu},\Sigma\right)=f_{\mathbf{x}^{n}}\left(\boldsymbol{\mu}|\Sigma\right)f_{\mathbf{x}^{n}}\left(\Sigma\right)$.
Define $A\stackrel{\text{def}}{=}f\left(\mathbf{x}_{n+1}|\mathbf{x}^{n}\right)=\int_{\Sigma>0}\int f\left(\mathbf{x}_{n+1}|\boldsymbol{\mu},\Sigma\right)f_{\mathbf{x}^{n}}\left(\boldsymbol{\mu},\Sigma\right)d\boldsymbol{\mu}d\Sigma$

\begin{align*}
A & =B\int_{\Sigma>0}\det\left(\Sigma\right)^{-\frac{n+M+2}{2}}e_{1}\left(\Sigma\right)e_{2}\left(\Sigma\right)d\Sigma
\end{align*}
where 
\begin{align*}
e_{1}\left(\Sigma\right) & =\mathrm{etr}\left\{ -\frac{1}{2}\left(\hat{\Sigma}_{n}+n\hat{\boldsymbol{\mu}}_{n}\hat{\boldsymbol{\mu}}_{n}^{T}+\mathbf{x}_{n+1}\mathbf{x}_{n+1}^{T}\right)\Sigma^{-1}\right\} \\
e_{2}\left(\Sigma\right) & =\int\exp\left\{ -\frac{n+1}{2}\left[\boldsymbol{\mu}^{T}\Sigma^{-1}\boldsymbol{\mu}-2\hat{\boldsymbol{\mu}}_{n+1}\Sigma^{-1}\boldsymbol{\mu}\right]\right\} d\boldsymbol{\mu}\\
 & =\sqrt{\frac{\left(2\pi\right)^{M}\det\left(\Sigma\right)}{\left(n+1\right)^{M}}}\exp\left\{ \frac{n+1}{2}\hat{\boldsymbol{\mu}}_{n+1}^{T}\Sigma^{-1}\hat{\boldsymbol{\mu}}_{n+1}\right\} \\
B & =\frac{\det\left(\hat{\Sigma}_{n}\right)^{\frac{n-1}{2}}}{\Gamma_{M}\left(\frac{n-1}{2}\right)}\frac{n^{\frac{M}{2}}}{2^{\frac{M\left(n-1\right)}{2}}\left(2\pi\right)^{M}}
\end{align*}
now since $\hat{\Sigma}_{n+1}=\hat{\Sigma}_{n}+n\hat{\boldsymbol{\mu}}_{n}\hat{\boldsymbol{\mu}}_{n}^{T}+\mathbf{x}_{n+1}\mathbf{x}_{n+1}^{T}-\left(n+1\right)\hat{\boldsymbol{\mu}}_{n+1}\hat{\boldsymbol{\mu}}_{n+1}^{T}$,
by defining $C\stackrel{\text{def}}{=}B\sqrt{\frac{\left(2\pi\right)^{M}}{\left(n+1\right)^{M}}}=\sqrt{\left(\frac{n}{n+1}\right)^{M}}\frac{\det\left(\hat{\Sigma}_{n}\right)^{\frac{n-1}{2}}}{\Gamma_{M}\left(\frac{n-1}{2}\right)}\frac{1}{2^{\frac{M\left(n-1\right)}{2}}\left(2\pi\right)^{\frac{M}{2}}}$
we can write 
\begin{align*}
A & =C\int_{\Sigma>0}\det\left(\Sigma\right)^{-\frac{n+M+1}{2}}\mathrm{etr}\left\{ -\frac{1}{2}\hat{\Sigma}_{n+1}\Sigma^{-1}\right\} d\Sigma\\
 & =C\int_{Y>0}\det\left(Y\right)^{\frac{n}{2}-\frac{M+1}{2}}\mathrm{etr}\left\{ -\frac{1}{2}\hat{\Sigma}_{n+1}Y\right\} dY\\
 & =C\frac{2^{\frac{Mn}{2}}}{\det\left(\hat{\Sigma}_{n+1}\right)^{\frac{n}{2}}}\int_{V>0}\det\left(V\right)^{\frac{n}{2}-\frac{M+1}{2}}\mathrm{etr}\left\{ -V\right\} dV\\
 & =\frac{1}{\pi^{\frac{M}{2}}}\sqrt{\left(\frac{n}{n+1}\right)^{M}}\frac{\det\left(\hat{\Sigma}_{n}\right)^{\frac{n-1}{2}}}{\det\left(\hat{\Sigma}_{n+1}\right)^{\frac{n}{2}}}\frac{\Gamma_{M}\left(\frac{n}{2}\right)}{\Gamma_{M}\left(\frac{n-1}{2}\right)}
\end{align*}

\end{document}